\documentclass[11pt,a4paper]{article}

\usepackage{amsmath,amssymb,amsfonts}
\usepackage{amsthm}
\usepackage{graphicx}
\usepackage{booktabs}
\usepackage{multirow}
\usepackage{xcolor}
\usepackage{natbib}

\usepackage{mathrsfs}
\usepackage{bm}
\usepackage{tabularx}
\usepackage{siunitx}
\usepackage{placeins}
\usepackage{enumitem}
\usepackage{algorithm}
\usepackage{algorithmicx}
\usepackage{algpseudocode}
\usepackage{listings}
\usepackage{textcomp}
\usepackage[title]{appendix}

\usepackage{url}
\usepackage{hyperref}
\hypersetup{
  colorlinks=true,
  linkcolor=blue,
  citecolor=blue,
  urlcolor=blue,
  pdftitle={Benchmarking Quantum Kernel Support Vector Machines Against Classical Baselines},
  pdfauthor={Kakavand, Siavash and Strohmeyer, Christoph and Schlotter, Michael},
  pdfkeywords={Quantum machine learning, Quantum kernels, SVM, Benchmarking}
}

\providecommand{\doi}[1]{}
\renewcommand{\doi}[1]{\href{https://doi.org/#1}{\texttt{doi:#1}}}

\usepackage[left=1in,right=1in,top=1in,bottom=1in]{geometry}
\usepackage{setspace}
\theoremstyle{plain}

\theoremstyle{definition}

\theoremstyle{remark}

\newcommand{\BA}{\mathrm{BA}}
\newcommand{\KTA}{\mathrm{KTA}}
\newcommand{\CoV}{\mathrm{CoV}}

\let\botrule\bottomrule
\newcommand{\tr}{\operatorname{Tr}}
\newcommand{\ket}[1]{|#1\rangle}
\newcommand{\bra}[1]{\langle#1|}
\newcommand{\braket}[2]{\langle#1|#2\rangle}
\newcommand{\ketbra}[2]{|#1\rangle\langle#2|}

\graphicspath{{figures/}}

\begin{document}

\title{Benchmarking Quantum Kernel Support Vector Machines Against Classical Baselines on Tabular Data:\\ A Rigorous Empirical Study with Hardware Validation}

\author{Siavash Kakavand\thanks{Corresponding author: siavash.kakavand@rwth-aachen.de}, Christoph Strohmeyer, Michael Schlotter \\ \small SHARE at FAU, Schaeffler Technologies AG \& Co.\ KG, Herzogenaurach 91074, Bavaria, Germany}

\date{April 2026}

\maketitle

\begin{abstract}
Quantum kernel methods have been proposed as a promising approach for leveraging near-term quantum computers for supervised learning, yet rigorous benchmarks against strong classical baselines remain scarce. We present a comprehensive empirical study of quantum kernel support vector machines (QSVMs) across nine binary classification datasets, four quantum feature maps, three classical kernels, and multiple noise models, totalling 970 experiments with strict nested cross-validation.

Our analysis spans four phases: (i)~statistical significance testing, revealing that none of 29 pairwise quantum--classical comparisons reach significance at $\alpha=0.05$; (ii)~learning curve analysis over six training fractions, showing steeper quantum slopes on six of eight datasets that nonetheless fail to close the gap to the best classical baseline; (iii)~hardware validation on IBM ibm\_fez (Heron~r2), demonstrating kernel fidelity $r \geq 0.976$ across six experiments; and (iv)~seed sensitivity analysis confirming reproducibility (mean CV 1.4\%).

A Kruskal--Wallis factorial analysis reveals that dataset choice dominates performance variance ($\varepsilon^2 = 0.73$), while kernel type accounts for only 9\%. Spectral analysis offers a mechanistic explanation: current quantum feature maps produce eigenspectra that are either too flat or too concentrated, missing the intermediate profile of the best classical kernel, the radial basis function (RBF). Quantum kernel training (QKT) via kernel-target alignment yields the single competitive result -- balanced accuracy 0.968 on breast cancer -- but with $\sim2{,}000\times$ computational overhead.

Our findings provide actionable guidelines for quantum kernel research. The complete benchmark suite is publicly available to facilitate reproduction and extension.

\end{abstract}


\section{Introduction}
\label{sec:introduction}

The prospect of achieving computational advantages through quantum computing has generated significant interest across machine learning~\citep{biamonte2017quantum, schuld2019quantum, havlicek2019supervised}. Among the most promising near-term approaches are \emph{quantum kernel methods}, which leverage parameterised quantum circuits to embed classical data into exponentially large Hilbert spaces, where inner products serve as kernel evaluations for classical support vector machines (SVMs)~\citep{rebentrost2014quantum, havlicek2019supervised, schuld2021supervised}. This framework is theoretically appealing: it inherits the well-understood generalisation guarantees of classical kernel methods~\citep{vapnik1998statistical, shawe-taylor2004kernel} while potentially accessing feature spaces that are classically intractable to compute under certain structural conditions~\citep{liu2021rigorous}.

Despite this theoretical promise, the practical utility of quantum kernels on real-world datasets remains an open question. Several theoretical works have identified conditions under which quantum kernels can provably outperform classical methods~\citep{liu2021rigorous, huang2021power}, but these constructions typically require specially crafted datasets or access to the data-generating process. On standard tabular datasets -- the primary domain for SVMs in practice -- the evidence for quantum advantage is mixed at best~\citep{huang2021power, kuebler2021inductive, thanasilp2024exponential}. Moreover, the exponential concentration of quantum kernels in high-dimensional feature spaces~\citep{thanasilp2024exponential} and the limited expressiveness of shallow quantum circuits~\citep{schuld2021supervised} raise fundamental concerns about the scalability of these methods.

Existing benchmarks of quantum kernel SVMs suffer from several methodological limitations. Many studies (i)~compare against weak classical baselines (e.g., linear SVMs only), (ii)~use simple hold-out validation rather than nested cross-validation (CV), risking optimistic bias~\citep{cawley2010over, varma2006bias}, (iii)~evaluate on a single or few datasets without statistical significance testing~\citep{demsar2006statistical}, or (iv)~rely entirely on simulation without hardware validation. The few studies that address some of these concerns~\citep{schnabel2025quantum} typically do not combine all elements into a single, comprehensive framework that also includes hardware execution on real quantum processors.

In this work, we present a rigorous, large-scale empirical study of quantum kernel SVMs that addresses all of the above limitations simultaneously. Our benchmark comprises 970 experiments across nine binary classification datasets, four quantum feature maps, three classical kernel baselines, and three backend types (ideal simulation, noisy simulation, IBM quantum hardware), evaluated using strict nested cross-validation (main benchmark: $5 \times 3$ folds; extended study: $5 \times 5$ folds). We structure our analysis into four complementary phases:

\begin{enumerate}[label=(\roman*)]
    \item \textbf{Statistical analysis} (Sect.~\ref{sec:results_statistical}): Paired Wilcoxon signed-rank tests across 29 quantum--classical comparisons, Kruskal--Wallis factorial analysis of seven experimental factors, and spectral analysis of 95 kernel matrices.
    
    \item \textbf{Learning curves} (Sect.~\ref{sec:results_learning}): Evaluation at six training fractions (10\%--100\%) to assess data efficiency and convergence behaviour across 720 data points.
    
    \item \textbf{Hardware validation} (Sect.~\ref{sec:results_hardware}): Six experiments on IBM ibm\_fez (Heron~r2 processor, 156~qubits) computing fidelity-based kernel matrices and comparing against cached simulation results.
    
    \item \textbf{Seed sensitivity} (Sect.~\ref{sec:results_seed}): Reproducibility analysis across 16 random seeds for 21 representative configurations, totalling 8\,400 SVM fits.
\end{enumerate}

The remainder of this paper is organised as follows. Section~\ref{sec:background} introduces the mathematical framework for quantum kernel methods. Section~\ref{sec:related_work} reviews related benchmarking studies. Section~\ref{sec:methodology} describes our benchmark methodology, including feature maps, kernels, and evaluation protocol. Section~\ref{sec:experimental_setup} details the experimental configuration. Section~\ref{sec:results} presents results across all four phases. Section~\ref{sec:discussion} discusses implications and limitations. Section~\ref{sec:conclusion} concludes with recommendations for the quantum kernel community.

\section{Background}
\label{sec:background}

\subsection{Support Vector Machines and Kernel Methods}
\label{sec:bg_svm}

A support vector machine (SVM)~\citep{cortes1995support, vapnik1998statistical} solves the binary classification problem by finding the maximum-margin hyperplane in a (possibly transformed) feature space. Given training data $\{(\bm{x}_i, y_i)\}_{i=1}^{N}$ with $\bm{x}_i \in \mathbb{R}^d$ and $y_i \in \{-1, +1\}$, the soft-margin SVM optimises:
\begin{equation}
\label{eq:svm_primal}
\min_{\bm{w}, b, \bm{\xi}} \frac{1}{2}\|\bm{w}\|^2 + C \sum_{i=1}^{N} \xi_i, \quad \text{s.t.} \quad y_i(\bm{w}^\top \phi(\bm{x}_i) + b) \geq 1 - \xi_i, \quad \xi_i \geq 0,
\end{equation}
where $\phi: \mathbb{R}^d \to \mathcal{H}$ maps inputs to a reproducing kernel Hilbert space (RKHS) $\mathcal{H}$, $C > 0$ controls the regularisation--margin trade-off, and $\xi_i \geq 0$ are slack variables measuring the margin violation of sample~$i$. The dual formulation~\citep{shawe-taylor2004kernel} depends on the data only through the \emph{kernel matrix}:
\begin{equation}
\label{eq:kernel_matrix}
K_{ij} = k(\bm{x}_i, \bm{x}_j) = \langle \phi(\bm{x}_i), \phi(\bm{x}_j) \rangle_{\mathcal{H}}.
\end{equation}
By Mercer's theorem, any continuous symmetric positive semi-definite (PSD) function $k: \mathcal{X} \times \mathcal{X} \to \mathbb{R}$ on a compact domain $\mathcal{X}$ corresponds to an inner product in some RKHS. Common classical kernels include the linear kernel $k(\bm{x}, \bm{z}) = \bm{x}^\top \bm{z}$, the radial basis function (RBF) kernel $k(\bm{x}, \bm{z}) = \exp(-\gamma \|\bm{x} - \bm{z}\|^2)$, and the polynomial kernel $k(\bm{x}, \bm{z}) = (\gamma \bm{x}^\top \bm{z} + r)^p$.

\subsection{Quantum Kernel Methods}
\label{sec:bg_quantum_kernels}

Quantum kernel methods~\citep{havlicek2019supervised, schuld2019quantum} replace the classical feature map $\phi$ with a quantum feature map $U(\bm{x})$ that encodes classical data into quantum states:
\begin{equation}
\label{eq:quantum_encoding}
\ket{\psi(\bm{x})} = U(\bm{x}) \ket{0}^{\otimes n},
\end{equation}
where $U(\bm{x})$ is a parameterised unitary acting on $n$ qubits, and $\ket{0}^{\otimes n}$ is the all-zero initial state. The quantum kernel is then defined as the fidelity between encoded states:
\begin{equation}
\label{eq:quantum_kernel}
k_Q(\bm{x}, \bm{z}) = |\braket{\psi(\bm{x})}{\psi(\bm{z})}|^2 = |\bra{0}^{\otimes n} U^\dagger(\bm{x}) U(\bm{z}) \ket{0}^{\otimes n}|^2.
\end{equation}
This fidelity kernel is symmetric, bounded in $[0, 1]$, satisfies $k_Q(\bm{x}, \bm{x}) = 1$, and is PSD by construction~\citep{schuld2021supervised}, since $k_Q(\bm{x}, \bm{z}) = |\braket{\psi(\bm{x})}{\psi(\bm{z})}|^2 = \tr(\rho_{\bm{x}} \rho_{\bm{z}})$ where $\rho_{\bm{x}} = \ketbra{\psi(\bm{x})}{\psi(\bm{x})}$. On a quantum computer, $k_Q(\bm{x}, \bm{z})$ is estimated via the \emph{compute-uncompute} circuit: apply $U(\bm{x})$, then $U^\dagger(\bm{z})$, and measure the probability of the all-zero bitstring.

The implicit quantum feature map $\phi_Q(\bm{x}) = \ket{\psi(\bm{x})}$ embeds data into a $2^n$-dimensional Hilbert space. For $n$ qubits, the feature space dimension grows exponentially, which may provide representational power beyond efficient classical simulation~\citep{havlicek2019supervised}. However, Schuld~\citep{schuld2021supervised} proved that all quantum models with a fixed encoding circuit are mathematically equivalent to classical kernel methods, implying that any potential advantage must come from the \emph{specific structure} of the quantum feature map rather than from quantum computation per se.

\subsection{Quantum Feature Maps}
\label{sec:bg_feature_maps}

The choice of quantum feature map $U(\bm{x})$ is critical for kernel quality. Feature maps differ in their entanglement structure, circuit depth, and data encoding strategy. We evaluate four families, detailed in Sect.~\ref{sec:methodology_feature_maps} and Appendix~\ref{app:feature_maps}:

\begin{itemize}
    \item \textbf{Rot2DoF}: Product-state encoding with $R_X$ and $R_Z$ rotations per qubit (no entanglement). Constant depth $= 12$ at $\text{reps} = 2$.
    \item \textbf{Belis}~\citep{belis2024quantum}: Controlled-NOT (CNOT) ladder entanglement with $R_X/R_Z$ data encoding. Depth scales linearly with $n$.
    \item \textbf{Sakhnenko10}~\citep{sakhnenko2022hybrid}: Ring-CNOT topology with double-feature $R_X$ encoding interleaved with fixed $R_Y(\pi/2)$ and $R_X(\pi/2)$ rotations per qubit, yielding a four-gate single-qubit block $R_X(x_1)\cdot R_Y(\pi/2)\cdot R_X(\pi/2)\cdot R_X(x_2)$ per qubit, where $x_1, x_2$ denote the two encoded features.
    \item \textbf{ZZFeatureMap}~\citep{havlicek2019supervised}: Hadamard preparation followed by pairwise $ZZ$-interaction data encoding.
\end{itemize}

\subsection{Trainable Quantum Kernels}
\label{sec:bg_qkt}

Quantum Kernel Training (QKT)~\citep{hubregtsen2022training} introduces learnable parameters $\bm{\theta}$ that scale the input features before encoding: $\bm{x} \mapsto \bm{\theta} \odot \bm{x}$. The parameters are optimised to maximise the \emph{centred kernel-target alignment} (KTA)~\citep{cristianini2001kernel, cortes2012algorithms}:
\begin{equation}
\label{eq:kta}
\KTA(\bm{\theta}) = \frac{\langle \tilde{K}_{\bm{\theta}}, \tilde{Y} \rangle_F}{\|\tilde{K}_{\bm{\theta}}\|_F \|\tilde{Y}\|_F},
\end{equation}
where $\tilde{K}_{\bm{\theta}}$ is the centred kernel matrix with parameters $\bm{\theta}$, $\tilde{Y} = \tilde{\bm{y}} \tilde{\bm{y}}^\top$ is the centred ideal kernel, $\langle \cdot, \cdot \rangle_F$ denotes the Frobenius inner product, and centring removes the kernel mean: $\tilde{K} = HKH$ with $H = I - \frac{1}{N}\bm{1}\bm{1}^\top$. Centring ensures that KTA measures alignment in the zero-mean subspace, making the metric invariant to constant shifts in kernel entries and thus a more robust objective for optimisation. The optimisation is performed using the efficient, gradient-based algorithm Limited-memory Broyden--Fletcher--Goldfarb--Shanno with Bound constraints (L-BFGS-B) with box constraints $\theta_j \in [0.01, 5.0]$. The lower bound prevents degenerate zero-scaling that would effectively remove a feature from the encoding, while the upper bound limits extreme amplification that causes rotation-angle wrapping and numerical instability in the kernel.

Crucially, to avoid data leakage, QKT parameters must be optimised \emph{per fold} on training data only, as the optimal $\bm{\theta}$ depends on the label distribution.

\subsection{Kernel Matrix Spectral Analysis}
\label{sec:bg_spectral}

The spectral properties of a kernel matrix $K$ provide insight into its suitability for SVM classification. We characterise kernels by:
\begin{itemize}
    \item \textbf{Effective rank ratio} (normalised spectral entropy): $\hat{r} = \exp\!\bigl(H(\bm{p})\bigr) / N$, where $H(\bm{p}) = -\sum_i p_i \ln p_i$ is the Shannon entropy of the normalised eigenvalue distribution $p_i = \lambda_i / \sum_j \lambda_j$ (with $\lambda_i > 0$), and $N$ is the matrix size. Values near 1 indicate a near-identity (flat) kernel; values near $1/N$ indicate rank-1 collapse.
    \item \textbf{Spectral concentration}: Fraction of total variance captured by the top-$k$ eigenvalues.
    \item \textbf{Diagonal dominance}: Ratio $\bar{K}_{\text{diag}} / \bar{K}_{\text{off-diag}}$. High values indicate a near-identity kernel where all sample pairs are projected to nearly orthogonal states.
\end{itemize}

An effective SVM kernel should have an intermediate spectral profile: sufficient eigenvalue concentration to provide discriminative structure for the decision boundary, but not so extreme as to collapse information (cf.\ the general spectral analysis framework in \citet{shawe-taylor2004kernel}). As we show in Sect.~\ref{sec:results_statistical}, current quantum feature maps tend toward one extreme or the other.

\section{Related Work}
\label{sec:related_work}

\paragraph{Theoretical landscapes}
The theoretical underpinnings of quantum kernel methods were established by Havl{\'i}{\v{c}}ek et al.~\citep{havlicek2019supervised}, who demonstrated that quantum computers can generate kernel functions corresponding to classically intractable feature maps, and by Schuld and Killoran~\citep{schuld2019quantum}, who formalised the connection between quantum circuits and kernel methods. Liu et al.~\citep{liu2021rigorous} proved a rigorous quantum speed-up for a carefully constructed classification problem, establishing that provable advantages exist in principle. However, Huang et al.~\citep{huang2021power} showed that the advantage depends critically on the relationship between the quantum feature map and the data distribution -- classical methods equipped with a \emph{prediction function} derived from the quantum model can match its performance when classical data is abundant.

\paragraph{Concentration and trainability}
K{\"u}bler et al.~\citep{kuebler2021inductive} analysed the inductive bias of quantum kernels, demonstrating that random quantum feature maps tend to produce kernels that are essentially random inner products in high-dimensional spaces, offering no useful structure for classification. Thanasilp et al.~\citep{thanasilp2024exponential} proved that quantum kernels exhibit \emph{exponential concentration}: as the number of qubits grows, all kernel values concentrate around a fixed value, rendering the kernel matrix uninformative. These results suggest that unstructured quantum feature maps cannot provide systematic advantages, and that problem-specific circuit design is essential.

\paragraph{Quantum kernel training}
To address the limitations of fixed feature maps, Hubregtsen et al.~\citep{hubregtsen2022training} proposed training quantum embedding kernels by optimising learnable parameters to maximise kernel-target alignment (KTA). Glick et al.~\citep{glick2024covariant} introduced covariant quantum kernels that exploit group structure in the data. Shaydulin and Wild~\citep{shaydulin2022importance} demonstrated the importance of kernel bandwidth tuning, showing that carefully scaled quantum kernels can significantly improve classification performance.

\paragraph{Benchmarking studies}
Several recent studies have benchmarked quantum kernels against classical methods. Peters et al.~\citep{peters2021machine} evaluated quantum kernel methods on high-dimensional cosmological data (67 features) using Google's Sycamore processor, demonstrating that kernel-based classification on real data is feasible on current hardware with careful attention to shot statistics. Belis et al.~\citep{belis2024quantum} benchmarked QSVMs on a trapped-ion device for high-energy physics classification, reporting competitive performance with specific feature map designs. Schnabel and Roth~\citep{schnabel2025quantum} performed a scrutiny-focused benchmark in the same journal, highlighting the importance of proper evaluation methodology and identifying common pitfalls in existing studies.

\paragraph{Our contribution in context}
Our work builds upon and extends these studies in several key dimensions, summarised in Table~\ref{tab:related_work}. Specifically, we combine: (i)~nine datasets spanning a range of sizes, dimensionalities, and class balances (vs.\ typically 2--5), (ii)~strict nested cross-validation throughout (vs.\ hold-out or simple $k$-fold), (iii)~comprehensive statistical testing with Wilcoxon, Friedman, and Kruskal--Wallis tests including effect sizes, (iv)~both fixed and trainable quantum kernels, (v)~hardware validation on IBM Quantum hardware alongside simulation, (vi)~spectral analysis providing mechanistic explanations, and (vii)~seed sensitivity analysis for reproducibility assessment. To our knowledge, no single prior study integrates all seven of these dimensions simultaneously.

\begin{table}[t]
\caption{Comparison of quantum kernel benchmarking studies. \checkmark~indicates inclusion; --~indicates absence. NCV: nested cross-validation; Stats: statistical tests; Spec.: spectral analysis; Seed: seed sensitivity analysis.}
\label{tab:related_work}
\setlength{\tabcolsep}{2.5pt}
\begin{tabular}{@{}lcccccccc@{}}
\toprule
\textbf{Study} & \textbf{Year} & \textbf{Data} & \textbf{NCV} & \textbf{Stats} & \textbf{QKT} & \textbf{HW} & \textbf{Spec.} & \textbf{Seed}\\
\midrule
Havl{\'i}{\v{c}}ek et al. & 2019 & 1 & -- & -- & -- & \checkmark & -- & --\\
Peters et al. & 2021 & 1 & -- & -- & -- & \checkmark & -- & --\\
Hubregtsen et al. & 2022 & 5 & -- & -- & \checkmark & -- & -- & --\\
Belis et al. & 2024 & 2 & -- & -- & -- & \checkmark & -- & --\\
Schnabel \& Roth & 2025 & $>$5 & \checkmark & \checkmark & -- & -- & -- & --\\
\midrule
\textbf{This work} & \textbf{2026} & \textbf{9} & \checkmark & \checkmark & \checkmark & \checkmark & \checkmark & \checkmark\\
\botrule
\end{tabular}
\end{table}

\section{Methodology}
\label{sec:methodology}

\subsection{Datasets}
\label{sec:methodology_datasets}

We evaluate on nine binary classification datasets from the UCI Machine Learning Repository~\citep{uci2017} and OpenML~\citep{vanschoren2014openml}, selected to span a range of sample sizes ($N \in [208, \num{4601}]$), dimensionalities ($d \in [3, 60]$), and class balances. Table~\ref{tab:datasets} summarises the dataset characteristics. While the majority of datasets originate from biomedical and physical-science domains, they cover distinct statistical profiles (feature counts, class ratios, separability) and are standard benchmarks in the classical SVM literature, enabling direct comparison with established baselines.

\begin{table}[t]
\caption{Dataset characteristics. $N$: post-cleaning sample count; $d$: original features; Class ratio: majority/minority class balance.}
\label{tab:datasets}
\begin{tabular}{@{}lrrrl@{}}
\toprule
\textbf{Dataset} & $N$ & $d$ & \textbf{Class Ratio} & \textbf{Domain}\\
\midrule
banknote          & \num{1372} & 4  & 56/44 & Image authentication\\
breast\_cancer    &   569  & 30 & 63/37 & Medical diagnosis\\
diabetes\_pima    &   768  & 8  & 65/35 & Medical diagnosis\\
haberman          &   306  & 3  & 74/26 & Survival analysis\\
heart\_disease    &   303  & 13 & 54/46 & Medical diagnosis\\
ionosphere        &   351  & 34 & 64/36 & Radar signal\\
parkinson\_489    &   489  & 22 & 75/25 & Speech analysis\\
sonar             &   208  & 60 & 53/47 & Signal classification\\
spambase          & \num{4601} & 57 & 61/39 & Text classification\\
\botrule
\end{tabular}
\end{table}

\subsection{Preprocessing and Dimensionality Reduction}
\label{sec:methodology_preprocessing}

Quantum feature maps require low-dimensional input, because embedding data into an excessively large Hilbert space often results in exponentially vanishing inner products, rendering the quantum kernel uninformative as the Gram matrix converges toward the identity matrix. This necessitates dimensionality reduction from the original feature space to $k \in \{3, 4, 6, 8, 10, 12, 14\}$ dimensions. The number of qubits depends on the encoding strategy: feature maps with \emph{double-feature encoding} (Rot2DoF, Belis, Sakhnenko10) embed two features per qubit via paired single-qubit rotations, requiring $n = \lceil k/2 \rceil$ qubits for $k$ features; ZZFeatureMap uses \emph{single-feature encoding} with one qubit per feature, requiring $n = k$ qubits. We employ three reduction strategies:

\begin{itemize}
    \item \textbf{PCA} (Principal Component Analysis): An unsupervised linear projection that retains the top-$k$ orthogonal principal components maximising explained variance~\citep{scikit-learn}, preceded by standard scaling to zero mean and unit variance.
    \item \textbf{NMF} (Non-negative Matrix Factorisation)~\citep{lee1999learning}: A parts-based decomposition that factorises the feature matrix into $k$ non-negative components, preserving additive structure. Preceded by min-max scaling to $[0,1]$ to enforce the non-negativity constraint.
    \item \textbf{Tree} (SelectFromModel): A supervised feature-selection method that ranks the original features by Gini importance from a fitted \texttt{DecisionTreeClassifier} and selects the top-$k$ features, preceded by standard scaling. Unlike PCA and NMF, this retains interpretable original features rather than constructing linear combinations.
\end{itemize}

All preprocessing -- scaling and dimensionality reduction -- is applied \emph{per outer CV fold}: the scaler and reducer are fit exclusively on the training split and then used to transform both training and test splits. This prevents information leakage from test-fold statistics into the feature transforms. For quantum experiments, the per-fold design requires computing a separate kernel matrix for each outer fold (rather than a single global $N \times N$ matrix), increasing the kernel budget by a factor equal to the number of outer folds. We accept this additional cost to ensure that absolute performance estimates are unbiased.

Reduced features are passed directly to the quantum feature-map circuits as rotation-gate angles (no further rescaling to a fixed range such as $[0,\pi]$ is applied).

\subsection{Quantum Feature Maps}
\label{sec:methodology_feature_maps}

We evaluate four quantum feature maps spanning different entanglement topologies and encoding strategies. All circuits use native-gate decomposition ($SX + RZ$) for hardware compatibility. The circuit depth is controlled by the \texttt{reps} parameter (default: $\text{reps} = 2$). We briefly describe each feature map below; full circuit diagrams and per-qubit-count properties are provided in Appendix~\ref{app:feature_maps} (Fig.~\ref{fig:circuits_all}).

\paragraph{Rot2DoF} A product-state encoding with no entanglement. Each qubit encodes two features via paired rotations: $U_{\text{rot}}(\bm{x}) = \prod_{r=1}^{R} \prod_{q=0}^{n-1} R_X(x_{2q})\, R_Z(x_{2q+1})$. Its constant depth ($= 12$ at $R=2$) makes it the shallowest circuit, and the absence of two-qubit gates gives it the highest hardware fidelity.

\paragraph{Belis} Proposed by \citet{belis2024quantum} for high-energy physics classification. Each repetition applies an encoding block~$S_x$ with $R_X(x_{2q})\, R_Z(x_{2q+1})$ per qubit, a CNOT ladder for nearest-neighbour entanglement, and a second encoding block~$S'_x$ with swapped rotation axes ($R_Z(x_{2q})\, R_X(x_{2q+1})$). Like Rot2DoF, it uses double-feature encoding ($\lceil k/2 \rceil$ qubits).

\paragraph{Sakhnenko10} Based on \citet{sakhnenko2022hybrid}. Applies double-feature $R_X$ encoding with fixed intermediate rotations ($R_Y(\pi/2)$, $R_X(\pi/2)$) and ring-topology CX gates for cyclic entanglement. The ring structure creates all-to-all correlations even at moderate depth.

\paragraph{ZZFeatureMap} The standard Qiskit feature map based on \citet{havlicek2019supervised}. Encodes data through pairwise $ZZ$ interactions: $U_{\text{ZZ}}(\bm{x}) = \prod_{r} [ \prod_{i<j} e^{i x_i x_j Z_i Z_j} \prod_{i} H_i ]$. Uses single-feature encoding ($k$ qubits), and the product $x_i x_j$ creates second-order feature interactions.

\bigskip\noindent Table~\ref{tab:feature_maps} summarises the properties of the decomposed circuits.

\begin{table}[t]
\caption{Quantum feature map circuit properties at $\text{reps} = 2$ after native-gate decomposition. Depths and gate counts are for $k = 8$ features. CX: two-qubit controlled-NOT gates. Params: number of unique data-encoding parameters ($= k$). Qubits: number of qubits for $k = 8$ features. Post-transpilation metrics for hardware experiments are reported in Appendix~\ref{app:feature_maps}.}
\label{tab:feature_maps}
\centering
\footnotesize
\begin{tabularx}{\linewidth}{@{}lrrrr>{\raggedright\arraybackslash}X>{\raggedright\arraybackslash}l@{}}
\toprule
\textbf{Feature Map} & \textbf{Qubits} & \textbf{Depth} & \textbf{CX} & \textbf{Params} & \textbf{Entanglement} & \textbf{Encoding}\\
\midrule
rot2dof     & 4  & 12 & 0  & $k$ & None (product state)   & $R_X, R_Z$ (2/qubit)\\
belis       & 4  & 29 & 6  & $k$  & CNOT ladder (linear)   & $R_X, R_Z$ (2/qubit)\\
sakhnenko10 & 4  & 48 & 8  & $k$  & Ring CNOT              & $R_X$ (2/qubit)\\
zzfm        & 8  & 67  & 112  & $k$  & $ZZ$ interaction        & $R_Z$ (1/qubit)\\
\botrule
\end{tabularx}
\end{table}

\subsection{Kernel Computation}
\label{sec:methodology_kernel}

We compute kernel matrices via three pathways:

\paragraph{Ideal (statevector)} Exact overlap $k_Q(\bm{x}, \bm{z}) = |\braket{\psi(\bm{x})}{\psi(\bm{z})}|^2$ computed via statevector simulation. Each data point is embedded once: the feature-map circuit $U(\bm{x})$ is applied to $\ket{0}^{\otimes n}$ and the resulting $2^n$-dimensional statevector is stored. Kernel entries are then computed as classical inner products $|\braket{\psi(\bm{x})}{\psi(\bm{z})}|^2$, requiring $O(N)$ circuit simulations plus $O(N^2)$ classical dot products for an $N$-sample dataset. This provides the noise-free reference kernel matrix; diagonal entries are~$1$ by definition.

\paragraph{Noisy (density matrix)} The kernel is computed as the Hilbert--Schmidt inner product $k_Q(\bm{x}, \bm{z}) = \tr[\rho(\bm{x}) \cdot \rho(\bm{z})]$, where $\rho(\bm{x})$ is the density matrix of the encoded state $U(\bm{x})\ket{0}^{\otimes n}$ evolved under a depolarising noise model with single-qubit error rate $p_{1q} = 10^{-3}$ and two-qubit error rate $p_{2q} = 10^{-2}$, representative of current superconducting hardware. For pure states this equals the fidelity $|\braket{\psi(\bm{x})}{\psi(\bm{z})}|^2$; for the mixed states arising from decoherence the Hilbert--Schmidt inner product remains a valid positive semi-definite kernel but is no longer identical to the Uhlmann fidelity~\citep{havlicek2019supervised}. The density-matrix trace formulation avoids the $O(N^2)$ circuit overhead of the compute-uncompute approach: each data point requires only \emph{one} circuit simulation to obtain $\rho(\bm{x})$, and kernel entries are computed as classical matrix inner products. This provides an $O(N)$ circuit advantage while faithfully modelling incoherent noise.

\paragraph{Hardware (IBM QPU)} Fidelity-based compute-uncompute circuits executed on IBM ibm\_fez (Heron~r2, 156~qubits)~\citep{ibm_quantum2024} using Qiskit Runtime SamplerV2. For each pair $(\bm{x}, \bm{z})$, the circuit $U(\bm{x})$ followed by $U^\dagger(\bm{z})$ is applied and measured. Circuits are transpiled at \texttt{optimization\_level=3} to the native gate set $\{rz, sx, \text{controlled-Z (CZ)}, x\}$, minimising circuit depth. Each kernel entry is estimated from 1\,024~shots as the empirical probability of the all-zero bitstring: $\hat{k}(\bm{x}, \bm{z}) = \text{count}(\bm{0}) / 1024$. No readout error mitigation is applied; results therefore reflect raw hardware fidelity.

\paragraph{Kernel post-processing} Simulation kernels (ideal and noisy) are positive semi-definite (PSD) by construction: ideal kernels are Gram matrices of statevector inner products, and noisy kernels are Hilbert--Schmidt inner products of positive operators. No explicit eigenvalue clipping is applied. Hardware kernel matrices, which can be slightly indefinite due to shot noise, are also used without PSD projection: scikit-learn's \texttt{SVC} accepts pre-computed kernels that may be indefinite, and retaining the raw matrices preserves the unfiltered hardware signal for fidelity analysis.

\subsection{Classical Baselines}
\label{sec:methodology_classical}

We compare against three classical SVM kernels:
\begin{itemize}
    \item \textbf{Linear}: $k(\bm{x}, \bm{z}) = \bm{x}^\top \bm{z}$.
    \item \textbf{RBF}: $k(\bm{x}, \bm{z}) = \exp(-\gamma \|\bm{x} - \bm{z}\|^2)$ with $\gamma = 1/(d \cdot \operatorname{Var}(X))$ (scikit-learn \texttt{gamma="scale"}).
    \item \textbf{Polynomial} (degree 3): $k(\bm{x}, \bm{z}) = (\gamma \bm{x}^\top \bm{z})^3$, i.e., the homogeneous variant ($r = 0$ in the general form of Eq.~\ref{eq:kernel_matrix}'s discussion; scikit-learn \texttt{SVC} default \texttt{coef0=0}).
\end{itemize}
Classical kernels are computed analytically using scikit-learn~\citep{scikit-learn}, providing exact baselines without shot noise or hardware artefacts.

\subsection{Evaluation Protocol}
\label{sec:methodology_evaluation}

\paragraph{Nested cross-validation} All experiments use stratified nested cross-validation~\citep{cawley2010over, varma2006bias}: the main benchmark (V1) uses a $5 \times 3$ design (5~outer, 3~inner folds) and the extended study (V2) uses $5 \times 5$ (5~outer, 5~inner folds). In both cases, the outer folds provide unbiased performance estimation and the inner folds select the $C$-hyperparameter from $C \in \{0.01, 0.1, 1, 10, 100\}$. For the Parkinson~489 dataset, group-aware splits ensure no speaker appears in both training and test folds. For each outer fold, preprocessing is fit on the training split, the training and test features are transformed, and the kernel matrices are computed from the transformed features. Inner CV operates on the training-fold kernel sub-matrix.

\paragraph{Primary metric} Balanced accuracy $\BA = \frac{1}{2}\left(\frac{\text{TP}}{\text{TP}+\text{FN}} + \frac{\text{TN}}{\text{TN}+\text{FP}}\right)$ is reported as the primary metric due to class imbalance in several datasets. We also compute F1-score, Matthews correlation coefficient (MCC), receiver operating characteristic area under the curve (ROC-AUC), and precision-recall AUC.

\paragraph{Statistical testing} We employ nonparametric rank-based tests, which make no distributional assumptions about the per-fold accuracy values:
\begin{itemize}
    \item \emph{Wilcoxon signed-rank test}~\citep{wilcoxon1945individual}: Two-sided paired comparison of per-fold BA arrays ($n = 5$) between quantum and classical kernels. This test is the nonparametric counterpart of the paired $t$-test and is appropriate for small paired samples.
    \item \emph{Friedman test}~\citep{friedman1937use}: Omnibus test for differences among multiple methods within each dataset, the nonparametric equivalent of repeated-measures ANOVA. When significant, Nemenyi post-hoc tests identify which pairs of methods differ.
\end{itemize}
The Wilcoxon signed-rank and Friedman tests are the recommended nonparametric methods for classifier comparison across multiple datasets~\citep{demsar2006statistical}. In addition, we perform a factorial screening via:
\begin{itemize}
    \item \emph{Kruskal--Wallis H-test}~\citep{kruskal1952use}: One-way analysis of seven experimental factors (dataset, reducer, $k$, kernel family, backend, QKT flag, group) on balanced accuracy across all 222 extended-study configurations, with epsilon-squared ($\varepsilon^2$) effect sizes. Unlike the paired tests above, Kruskal--Wallis treats the experimental configurations as independent observations and identifies which design factors most influence performance -- the nonparametric equivalent of one-way ANOVA.
\end{itemize}
Within the Phase~4 seed sensitivity analysis, Bonferroni correction is applied across the seven per-dataset comparisons ($\alpha_{\text{corr}} = 0.05/7$).

\paragraph{Reproducibility} Deterministic seeding (seed~$= 42$) is used throughout. All experiments log the Python environment, library versions, git commit hash, and random states. Kernel matrices are cached as content-addressable NumPy compressed archive (\texttt{.npz}) files (xxhash-based keys) for exact reproducibility.

\section{Experimental Setup}
\label{sec:experimental_setup}

The benchmark is organised into two experimental campaigns -- a main benchmark and a targeted extended study -- plus four analysis phases, totalling 970 kernel evaluations and 8\,400 additional SVM fits.

\subsection{Main Benchmark (V1)}
\label{sec:setup_v1}

This campaign comprises 748 experiments across five systematic sweeps:

\begin{itemize}
    \item \textbf{Main comparison:} 8 datasets $\times$ 7 kernels (4 quantum + linear + rbf + poly) at default settings ($k = d_{\max}$, $\text{reps} = 2$, no noise).
    \item \textbf{Reducer ablation:} PCA vs.\ NMF vs.\ tree-based selection on representative dataset--kernel pairs.
    \item \textbf{Qubit scaling:} $k \in \{3, 4, 6, 8\}$ features with both PCA and NMF.
    \item \textbf{Circuit depth:} $\text{reps} \in \{1, 2, 3\}$ for all quantum feature maps.
    \item \textbf{Noise sensitivity:} Depolarising noise at $p_{1q} \in \{0, 5 \times 10^{-4}, 10^{-3}, 5 \times 10^{-3}\}$ using density-matrix simulation.
\end{itemize}

All experiments use $5 \times 3$ nested CV, producing 588 cached kernel matrices and 748 result records.

\subsection{Extended Study (V2)}
\label{sec:setup_v2}

Motivated by V1 findings, the extended study adds 222 targeted experiments:

\begin{itemize}
    \item \textbf{Higher-$k$:} $k \in \{10, 12, 14\}$ for breast\_cancer, ionosphere, and sonar with NMF and tree reducers. Tests whether more qubits improve quantum kernel performance.
    \item \textbf{Larger datasets:} spambase ($n = 1{,}000$ stratified subsample) and diabetes\_pima at $k \in \{4, 6, 8, 10\}$. Tests scalability to larger problems.
    \item \textbf{Quantum Kernel Training (QKT):} 28 QKT experiments across 7 datasets with per-fold theta optimisation via L-BFGS-B. Tests whether learnable feature scaling can close the quantum--classical gap.
\end{itemize}

The extended study uses $5 \times 5$ nested CV (more inner folds), generating 95 additional kernel matrices. QKT experiments each require 50--170 iterations of kernel recomputation per fold, resulting in $\sim2{,}000\times$ computational overhead vs.\ classical baselines.

\subsection{Analysis Phases}
\label{sec:setup_v3}

The analysis phases operate on existing data from both campaigns without generating new kernel matrices:

\begin{itemize}
    \item \textbf{Phase 1} (Statistical analysis): 29 Wilcoxon tests, 9 Friedman tests, 7-factor Kruskal--Wallis, QKT stability analysis (28 configs), kernel eigenvalue decomposition (95 matrices), computational cost analysis, and dataset suitability criterion.
    
    \item \textbf{Phase 2} (Learning curves): 720 data points from 24 representative configurations (8 datasets $\times$ 3 categories) at 6 training fractions (10\%, 20\%, 30\%, 50\%, 70\%, 100\%), using precomputed kernel sub-matrices.
    
    \item \textbf{Phase 3} (Hardware validation): 6 experiments on IBM ibm\_fez (Heron~r2, 156 qubits)~\citep{ibm_quantum2024}, designed to answer three research questions. \textbf{Q1} (Headline fidelity): Does the hardware kernel faithfully reproduce ideal simulation? \textbf{Q2} (Entanglement effect): How does entangling (belis) vs.\ product-state (rot2dof) encoding affect hardware performance? \textbf{Q3} (Cross-dataset generalisation): Does kernel fidelity hold across multiple datasets? Each experiment computes a $60 \times 60$ kernel matrix on a stratified subsample using 1\,830 compute-uncompute circuits at 1\,024 shots. Total: $\sim$11\,000 circuits. Table~\ref{tab:hw_experiments} details the experimental matrix.
    
    \item \textbf{Phase 4} (Seed sensitivity): 21 configs $\times$ 16 seeds $\times$ 5 folds $\times$ 5 inner folds $= 8{,}400$ SVM fits. The kernel matrix is held fixed; only the \texttt{StratifiedKFold random\_state} varies.
\end{itemize}

\begin{table}[t]
\caption{Hardware experiments on IBM ibm\_fez (Heron~r2). All use 1\,024~shots, $k = 6{-}10$ features, and $n_{\text{sub}} = 60$ stratified subsample.}
\label{tab:hw_experiments}
\centering
\small
\begin{tabular}{@{}llllrrr@{}}
\toprule
\textbf{ID} & \textbf{Dataset} & \textbf{Feature Map} & \textbf{Reducer} & $k$ & \textbf{Depth} & \textbf{CZ Gates}\\
\midrule
HW-01 & breast\_cancer & belis     & NMF & 10 & 74 & 16\\
HW-02 & breast\_cancer & rot2dof   & NMF & 10 & 22 & 0\\
HW-03 & sonar          & belis     & PCA & 6  & 62 & 8\\
HW-04 & sonar          & rot2dof   & PCA & 6  & 22 & 0\\
HW-05 & diabetes\_pima & belis     & NMF & 6  & 62 & 8\\
HW-06 & diabetes\_pima & rot2dof   & NMF & 6  & 22 & 0\\
\botrule
\end{tabular}
\end{table}

\subsection{Implementation}
\label{sec:setup_implementation}

The benchmark suite is implemented in Python~3.10+ using Qiskit~2.1+~\citep{javadi2024quantum} for quantum circuit construction and simulation, and scikit-learn~\citep{scikit-learn} for classical SVMs. The complete codebase comprises $\sim$18\,700 non-blank lines of Python across 22 modules and 51 scripts, with 132 unit tests achieving full pass status. All quantum circuits use native-gate ($SX + RZ$) decomposition. Hardware experiments use Qiskit Runtime SamplerV2 in Job mode (IBM Quantum Open Plan, no sessions) with \texttt{optimization\_level=3} transpilation. No readout error mitigation is applied; results reflect raw hardware fidelity.

Computation was performed on a standard workstation (11th~Gen Intel Core~i7-1185G7, 3.00~GHz, 16~GB RAM) for simulation, and IBM ibm\_fez (156-qubit Heron~r2 processor) for hardware experiments. Hardware wall times range from 11 to 74~minutes per experiment (dominated by queue wait times). The complete simulation benchmark requires approximately 48~hours of compute time.

\section{Results}
\label{sec:results}

We present results across four analysis phases: statistical analysis (Sect.~\ref{sec:results_statistical}), learning curves (Sect.~\ref{sec:results_learning}), hardware validation (Sect.~\ref{sec:results_hardware}), and seed sensitivity (Sect.~\ref{sec:results_seed}).

\subsection{Main Benchmark Overview}
\label{sec:results_main}

Table~\ref{tab:main_results} summarises the best-performing kernel for each dataset across both campaigns. The classical linear kernel achieves the highest mean balanced accuracy across datasets ($\overline{\BA} = 0.830$), followed by RBF ($0.825$) and polynomial ($0.807$). The best quantum kernel, rot2dof, achieves $\overline{\BA} = 0.649$ -- an 18.1 percentage point (pp) gap to linear. With QKT optimisation, the best quantum configuration (belis/tree/k=8 on breast\_cancer) reaches $\BA = 0.968$, narrowly approaching the classical ceiling.

\begin{table}[t]
\caption{Best balanced accuracy (BA) per dataset. Q-Ideal: best quantum kernel (ideal simulation); Q-QKT: best quantum kernel with QKT; Classical: best classical baseline. $\Delta$: Q-Ideal $-$ Classical.}
\label{tab:main_results}
\centering
\footnotesize
\begin{tabular}{@{}lccccr@{}}
\toprule
\textbf{Dataset} & \textbf{Classical} & \textbf{Kernel} & \textbf{Q-Ideal} & \textbf{Q-QKT} & $\Delta$\\
\midrule
haberman        & 0.549 & rbf    & \textbf{0.581} & 0.586 & $+$3.2\\
spambase        & 0.903 & rbf    & 0.888          & ---   & $-$1.6\\
ionosphere      & 0.935 & rbf    & 0.905          & 0.911 & $-$3.1\\
sonar           & 0.872 & rbf    & 0.813          & 0.748 & $-$5.9\\
diabetes\_pima  & 0.729 & linear & 0.669          & 0.676 & $-$6.0\\
breast\_cancer  & 0.976 & rbf    & 0.913          & \textbf{0.968} & $-$6.3\\
banknote        & 1.000 & rbf    & 0.921          & 0.871 & $-$7.9\\
parkinson\_489  & 0.829 & rbf    & 0.746          & ---   & $-$8.3\\
heart\_disease  & 0.844 & poly3  & 0.724          & 0.804 & $-$12.0\\
\botrule
\end{tabular}
\end{table}

The sole dataset showing quantum favourability is \emph{haberman} ($\Delta = +3.2$~pp), the smallest and hardest dataset in our suite with classical baselines at $\BA \approx 0.55$. On all other datasets, classical kernels dominate by 1.6--12.0~pp.

\begin{figure}[t]
    \centering
    \includegraphics[width=\columnwidth]{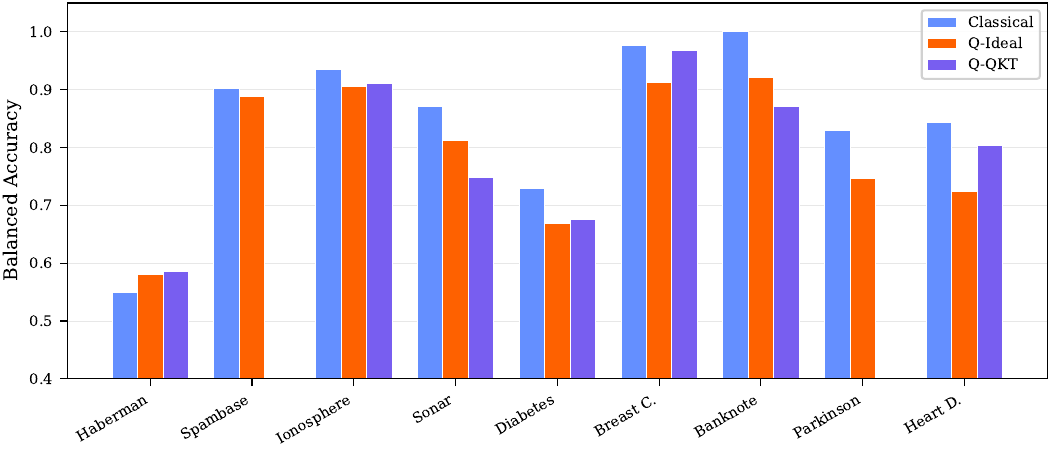}
    \caption{Balanced accuracy comparison across all nine datasets. For each dataset, bars show the best classical kernel (blue), best quantum ideal kernel (orange), and best quantum kernel with QKT (purple, where available). Classical kernels dominate on 8/9 datasets; haberman is the sole exception.}
    \label{fig:accuracy_comparison}
\end{figure}

\subsection{Statistical Significance and Factor Analysis}
\label{sec:results_statistical}

\paragraph{Zero significant quantum advantages}
Table~\ref{tab:wilcoxon} reports paired Wilcoxon signed-rank tests comparing the best quantum (ideal) kernel against the best classical kernel per dataset, using per-fold balanced accuracy arrays ($n = 5$ folds). None of the 29 pairwise comparisons reach statistical significance at $\alpha = 0.05$. The minimum achievable two-sided $p$-value for the exact Wilcoxon signed-rank test with $n = 5$ paired observations is $p = 0.0625$, which places a fundamental floor on statistical power. Several comparisons reach this floor (e.g., breast\_cancer, banknote, diabetes\_pima at $p = 0.0625$), suggesting that larger fold counts might reveal significance -- but in the \emph{wrong} direction for quantum advocates, as classical kernels outperform on 8/9 datasets.

\begin{table}[t]
\caption{Wilcoxon signed-rank tests: best quantum (ideal) vs.\ best classical per dataset. BA: balanced accuracy. $\Delta$: quantum $-$ classical BA. All tests non-significant at $\alpha = 0.05$.}
\label{tab:wilcoxon}
\centering
\small
\begin{tabular}{@{}llcrc@{}}
\toprule
\textbf{Dataset} & \textbf{Best Quantum} & \textbf{Best Classical} & $\Delta$ & $p$\\
\midrule
haberman        & sakhnenko10 & rbf    & $+$0.032 & 0.3125\\
spambase        & belis       & rbf    & $-$0.016 & 0.0625\\
ionosphere      & belis       & rbf    & $-$0.031 & 0.0625\\
sonar           & rot2dof     & rbf    & $-$0.059 & 0.1250\\
diabetes\_pima  & rot2dof     & linear & $-$0.060 & 0.0625\\
breast\_cancer  & rot2dof     & rbf    & $-$0.063 & 0.0625\\
banknote        & rot2dof     & rbf    & $-$0.079 & 0.0625\\
parkinson\_489  & rot2dof     & rbf    & $-$0.083 & 0.1250\\
heart\_disease  & rot2dof     & poly3  & $-$0.120 & 0.0625\\
\botrule
\end{tabular}
\end{table}

Friedman tests within each dataset confirm significant omnibus differences among methods ($p < 0.001$ for all 9 datasets), establishing that the kernel methods are not exchangeable. The lack of Wilcoxon significance is thus a sample-size limitation ($n = 5$), not an absence of performance differences.

\paragraph{Factorial analysis}
A Kruskal--Wallis analysis examines seven experimental factors on balanced accuracy across all 222 extended-study experiments. Dataset choice dominates with $\varepsilon^2 = 0.73$ ($p < 10^{-31}$), explaining 73\% of rank variance -- $8\times$ larger than kernel family ($\varepsilon^2 = 0.089$). Feature dimensionality ($k$) is the second most important factor ($\varepsilon^2 = 0.278$). Notably, the reducer (NMF vs.\ tree) and QKT flag are \emph{not} statistically significant, suggesting that these design choices matter less than the dataset itself.

\begin{figure}[t]
    \centering
    \includegraphics[width=0.85\columnwidth]{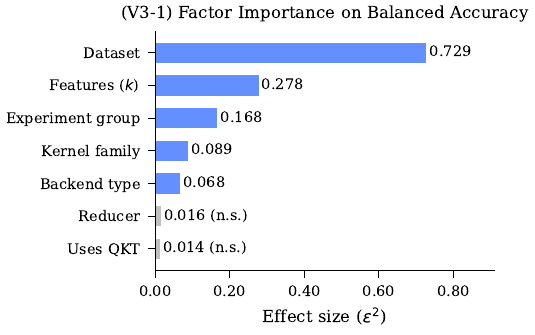}
    \caption{Kruskal--Wallis epsilon-squared ($\varepsilon^2$) effect sizes for seven experimental factors. Dataset choice dominates performance variance at $\varepsilon^2 = 0.73$, while kernel family contributes only $\varepsilon^2 = 0.089$.}
    \label{fig:factor_importance}
\end{figure}

\paragraph{Spectral analysis: The Goldilocks hypothesis}
Eigenvalue decomposition of 95 kernel matrices reveals a structural explanation for quantum kernel underperformance. Table~\ref{tab:spectral} shows that quantum feature maps produce extreme eigenspectra:

\begin{itemize}
    \item \textbf{Belis} kernels are \emph{near-uniform}: effective rank ratio 0.40--0.74, top-1 eigenvalue explains only 3--6\% of variance. The kernel matrix is approximately identity-like, providing minimal geometric structure for the SVM decision boundary.
    \item \textbf{Rot2dof} kernels are \emph{extremely concentrated}: top-1 eigenvalue captures 52--57\%, effective rank ratio 0.01--0.02. Information is collapsed onto essentially one principal component.
    \item \textbf{RBF} kernels occupy an intermediate ``\emph{Goldilocks zone}'': top-1 at 31--35\%, effective rank 0.06--0.07. This provides sufficient structure for discrimination while preserving multi-dimensional information.
\end{itemize}

\begin{table}[t]
\caption{Kernel matrix spectral properties across extended-study configurations ($n = 95$ kernel matrices, 3 datasets). Effective rank ratio normalised by matrix size $N$.}
\label{tab:spectral}
\centering
\small
\begin{tabular}{@{}lcccc@{}}
\toprule
\textbf{Kernel} & \textbf{Eff.\ Rank Ratio} & \textbf{Top-1 Var.} & \textbf{Top-5 Var.} & \textbf{Diag.\ Dom.}\\
\midrule
belis (quantum)   & 0.40--0.74 & 3--6\%   & 7--13\%  & 18--39\\
rot2dof (quantum) & 0.01--0.02 & 52--57\% & 66--72\% & 1.9--2.2\\
linear (classical)& 0.01       & 27--29\% & 76--80\% & 568\\
poly3 (classical) & 0.03       & 22--26\% & 62--63\% & 154--168\\
rbf (classical)   & 0.06--0.07 & 31--35\% & 52--54\% & 3.7--3.8\\
\botrule
\end{tabular}
\end{table}

Neither quantum kernel achieves the spectral profile of RBF, which quantitatively explains their consistent underperformance. This ``\emph{Goldilocks hypothesis}'' -- that an ideal kernel must be neither too flat nor too concentrated -- offers a constructive design criterion for future quantum feature maps.

\begin{figure}[t]
    \centering
    \includegraphics[width=\columnwidth]{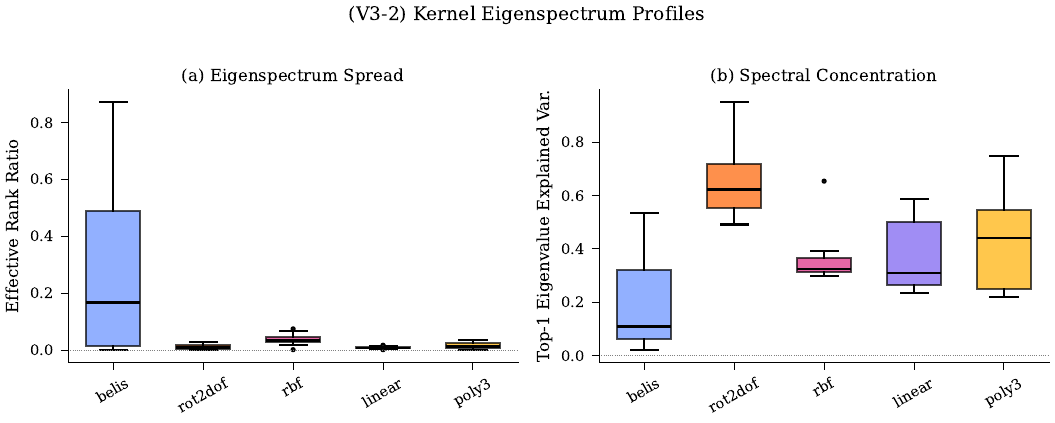}
    \caption{Kernel eigenspectra comparison across all extended-study configurations (3 datasets, $k \in \{4, 6, 8, 10, 12, 14\}$, NMF and tree reducers). Belis produces a near-uniform spectrum (too flat), rot2dof is heavily concentrated in the first eigenvalue (too peaked), while RBF occupies an intermediate ``Goldilocks zone'' that enables the best SVM performance.}
    \label{fig:eigenspectrum}
\end{figure}

\paragraph{Computational cost}
Quantum kernel computation via statevector simulation costs a mean of 3.07~s per kernel matrix, versus 0.31~s for classical kernels -- a $10\times$ overhead. QKT adds a further $\sim207\times$ factor (50--170 iterations), totalling $\sim2{,}060\times$ overhead (Fig.~\ref{fig:cost_comparison} in Appendix~\ref{app:tables}). Given the comparable or inferior accuracy, the cost--benefit ratio is strongly unfavourable for quantum kernels on the tested datasets.

\subsection{Learning Curve Analysis}
\label{sec:results_learning}

We evaluate 24 representative configurations (8 datasets $\times$ 3 categories) at six training fractions to assess data efficiency. Slopes are computed via ordinary least-squares (OLS) regression of balanced accuracy on $\log n_{\text{train}}$ across six fractions. Learning curves and full slope data are reported in Fig.~\ref{fig:learning_curves} and Table~\ref{tab:slopes} (Appendix~\ref{app:tables}).

\paragraph{Comparable learning rates, persistent classical lead}
On 6 of 8 datasets, the quantum--classical accuracy gap narrows slightly as training size increases, with quantum kernels exhibiting steeper learning slopes on 6/8 datasets (Table~\ref{tab:slopes}). The mean slope across quantum ideal configurations is 0.032 (BA per log-unit of $n_{\text{train}}$), matching the classical mean of 0.032. However, since quantum kernels start from a lower baseline, comparable learning rates do not translate into competitive final accuracy: classical baselines maintain or widen their lead at full training size on 8 of 9 datasets.

\paragraph{Haberman: the sole quantum success}
Only on haberman does the quantum kernel surpass the representative classical configuration at full training size ($\BA = 0.581$ vs.\ $0.534$, gap narrows from $-0.002$ to $+0.046$, where the learning-curve classical representative uses a fixed config rather than the best overall classical BA of 0.549), consistent with our Phase~1 finding that haberman is the sole dataset with marginal quantum favourability (Table~\ref{tab:main_results}).

\paragraph{Heart disease: quantum divergence}
On heart\_disease, quantum starts slightly ahead at 10\% training data ($+0.8$~pp) but falls far behind at 100\% ($-11.0$~pp), suggesting the quantum kernel's limited spectral structure prevents it from exploiting additional data.

\subsection{Hardware Validation on IBM Quantum}
\label{sec:results_hardware}

Six experiments on IBM ibm\_fez validate that quantum kernels computed on real hardware closely reproduce ideal simulation. Table~\ref{tab:hw_results} summarises kernel fidelity metrics and downstream SVM performance.

\begin{figure}[t]
    \centering
    \includegraphics[width=\columnwidth]{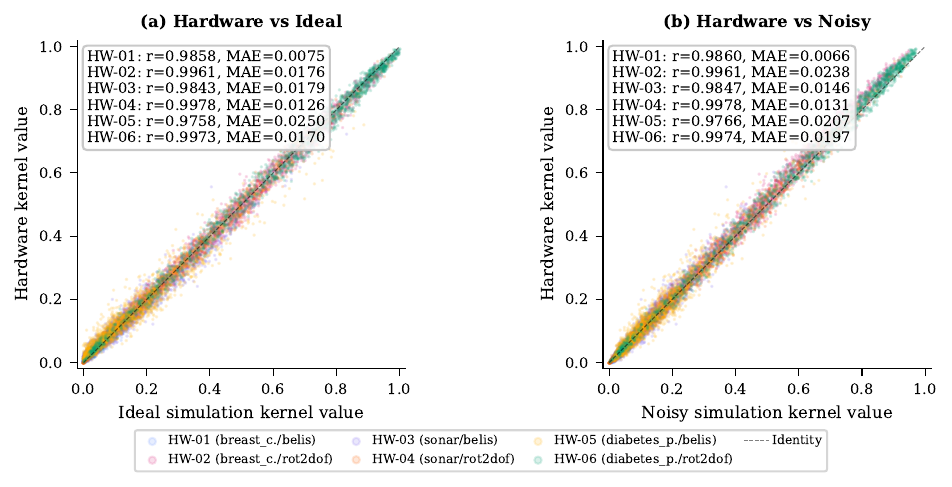}
    \caption{Hardware vs.\ simulation kernel entries for six experiments on IBM ibm\_fez (Heron~r2): (a)~hardware vs.\ ideal statevector simulation, (b)~hardware vs.\ noisy depolarising simulation. Each point represents one upper-triangular off-diagonal kernel entry; colours distinguish individual experiments (see legend). Pearson $r \geq 0.976$ for both ideal and noisy comparisons across all experiments, validating that both statevector and depolarising-noise simulations are faithful proxies for hardware kernel matrices.}
    \label{fig:hw_vs_sim}
\end{figure}

\begin{table}[t]
\caption{Hardware validation results on IBM ibm\_fez (Heron~r2). FM: feature map. HW: hardware. $r$: Pearson correlation between hardware and ideal kernel matrices (off-diagonal entries). BA: downstream 5-fold SVM balanced accuracy on $n=60$ subsample. $\Delta$: hardware $-$ ideal.}
\label{tab:hw_results}
\centering
\footnotesize
\begin{tabular}{@{}lllcccr@{}}
\toprule
\textbf{ID} & \textbf{Dataset} & \textbf{FM} & \textbf{Pearson~$r$} & \textbf{MAE} & \textbf{BA (HW)} & $\Delta$~pp\\
\midrule
HW-01 & breast\_cancer & belis   & 0.986 & 0.008 & 0.473 & $+$1.2\\
HW-02 & breast\_cancer & rot2dof & 0.996 & 0.018 & 0.760 & $+$7.5\\
HW-03 & sonar          & belis   & 0.984 & 0.018 & 0.550 & $+$3.3\\
HW-04 & sonar          & rot2dof & 0.998 & 0.013 & 0.544 & $-$2.6\\
HW-05 & diabetes\_pima & belis   & 0.976 & 0.025 & 0.511 & $-$5.5\\
HW-06 & diabetes\_pima & rot2dof & 0.997 & 0.017 & 0.538 & $+$5.2\\
\midrule
\multicolumn{3}{@{}l}{\textbf{Mean}} & \textbf{0.990} & \textbf{0.016} & --- & $+$1.5\\
\botrule
\end{tabular}
\end{table}

\paragraph{Kernel fidelity (Q1)}
Across all six experiments, the Pearson correlation between hardware and ideal kernel matrices satisfies $r \geq 0.976$ (exact minimum $r = 0.9758$ for HW-05, rounded to three decimal places in Table~\ref{tab:hw_results}), with a mean of $r = 0.990$. Mean absolute error (MAE) ranges from 0.008 to 0.025 per kernel entry. This validates statevector simulation as a faithful proxy for hardware at the circuit scales tested (3--5 qubits, depth 22--74).

\paragraph{Noise as regulariser}
Hardware kernels yield marginally higher BA than ideal simulation in 4 of 6 experiments (mean $\Delta = +1.5$~pp). While not statistically significant at $n = 60$, the consistent direction across experiments is noteworthy; we discuss the theoretical interpretation in Sect.~\ref{sec:discussion_hardware}.

\paragraph{Entanglement impact (Q2)}
Comparing feature map pairs within each dataset: on breast\_cancer, rot2dof outperforms belis by 28.7~pp ($\BA = 0.760$ vs.\ $0.473$); on sonar, the difference is negligible ($-0.6$~pp); on diabetes\_pima, rot2dof leads by $+2.7$~pp. The entanglement effect is thus \emph{dataset-dependent}. However, rot2dof consistently achieves higher kernel fidelity (mean $r = 0.997$ vs.\ $0.982$ for belis), attributable to its shallower hardware circuits (depth~22 vs.\ 62--74, zero CZ gates).

\paragraph{Cross-dataset generalisation (Q3)}
Kernel fidelity $r \geq 0.976$ across all three datasets confirms that hardware noise does not selectively degrade specific data distributions. The validation generalises from breast\_cancer ($k=10$, 5 qubits) to sonar ($k=6$, 3 qubits) and diabetes\_pima ($k=6$, 3 qubits).

\subsection{Seed Sensitivity Analysis}
\label{sec:results_seed}

Table~\ref{tab:seed_sensitivity} reports the stability of balanced accuracy across 16 random seeds for 21 representative configurations (7 datasets $\times$ 3 categories).

\begin{table}[t]
\caption{Seed sensitivity: coefficient of variation (CoV) of mean balanced accuracy (BA) across 16 seeds. ``QI Wins'': number of seeds where quantum ideal BA exceeds classical BA.}
\label{tab:seed_sensitivity}
\centering
\footnotesize
\begin{tabular}{@{}lccccrc@{}}
\toprule
\textbf{Dataset} & \textbf{CoV} & \textbf{CoV} & \textbf{CoV} & \textbf{QI Wins} & \textbf{Mean QI} & \textbf{Wilcoxon}\\
 & \textbf{Classical} & \textbf{Q-Ideal} & \textbf{Q-Noisy} & \textbf{/16} & \textbf{Adv.} & $p$\\
\midrule
banknote        & 0.000 & 0.002 & 0.004 & 0/16  & $-$0.084 & $<0.001$\\
breast\_cancer  & 0.005 & 0.007 & 0.007 & 0/16  & $-$0.069 & $<0.001$\\
diabetes\_pima  & 0.007 & 0.017 & 0.016 & 0/16  & $-$0.053 & $<0.001$\\
\textbf{haberman} & \textbf{0.033} & \textbf{0.026} & \textbf{0.024} & \textbf{14/16} & $\mathbf{+0.019}$ & $\mathbf{0.004}$\\
heart\_disease  & 0.019 & 0.018 & 0.014 & 0/16  & $-$0.072 & $<0.001$\\
ionosphere      & 0.008 & 0.008 & 0.008 & 0/16  & $-$0.055 & $<0.001$\\
sonar           & 0.018 & 0.028 & 0.024 & 0/16  & $-$0.081 & $<0.001$\\
\midrule
\textbf{Mean}   & \textbf{0.013} & \textbf{0.015} & \textbf{0.014} & --- & --- & ---\\
\botrule
\end{tabular}
\end{table}

\paragraph{High reproducibility}
The mean CoV across all configurations is 1.4\%, with maximum BA range 0.065 (heart\_disease, classical). The reference seed (42) is not an outlier -- results are robust to fold randomisation.

\paragraph{Resolving Phase~1 ambiguity}
With 16 paired observations (vs.\ 5 in Phase~1), Wilcoxon tests now achieve sufficient statistical power: all 7 datasets show \emph{significant} differences at $\alpha = 0.05$. On 6/7 datasets, classical kernels are significantly better ($p < 0.001$). On haberman, quantum kernels are significantly better ($p = 0.004$), winning on 14/16 seeds. The Phase~1 result of ``0/29 significant'' was entirely a sample-size limitation.

\paragraph{Haberman advantage is robust}
The quantum advantage on haberman ($+1.9$~pp mean, $p = 0.004$, 87.5\% win rate) persists across random seed perturbation, confirming it is not a statistical artefact of the reference seed.

\section{Discussion}
\label{sec:discussion}

\subsection{Why Quantum Kernels Underperform: A Structural Explanation}
\label{sec:discussion_spectral}

The spectral analysis in Sect.~\ref{sec:results_statistical} reveals that current quantum feature maps produce kernel matrices with extreme eigenspectra -- either near-identity (belis) or near-rank-1 (rot2dof) -- failing to provide the intermediate spectral structure that enables effective SVM decision boundaries.

The RBF kernel, by contrast, occupies a ``Goldilocks zone'' with moderate spectral concentration (effective rank ratio 0.06--0.07). This is no coincidence: the RBF bandwidth $\gamma$ is specifically tuned (via scikit-learn's auto-scaling) to match the data distribution, while quantum feature maps encode data through a fixed nonlinear transformation without such adaptive scaling.

\paragraph{Implications for quantum feature map design}
This analysis suggests that the quantum kernel community should prioritise feature maps with \emph{tunable spectral properties}. QKT~\citep{hubregtsen2022training} is one approach, but it is computationally expensive ($\sim2{,}000\times$ overhead). Alternative strategies include:
\begin{itemize}
    \item \emph{Bandwidth-aware encoding}: Introducing a learnable or data-dependent rescaling of features to $[0, \pi]$ before encoding, analogous to RBF~$\gamma$ tuning.
    \item \emph{Covariant kernels}: Leveraging data symmetries to construct structurally matched feature maps~\citep{glick2024covariant}.
    \item \emph{Spectral pre-screening}: Computing the effective rank ratio of a candidate quantum kernel \emph{before} running the full SVM pipeline, rejecting kernels with extreme profiles.
\end{itemize}

\subsection{Dataset Dominance and Benchmarking Practice}
\label{sec:discussion_datasets}

The finding that dataset choice explains 73\% of performance variance (Sect.~\ref{sec:results_statistical}) has important implications for the QML benchmarking community. It suggests that:
\begin{enumerate}
    \item Performance differences between quantum feature maps on a \emph{single} dataset are largely uninformative -- the dataset itself, not the kernel, drives the result.
    \item Future benchmarks should prioritise \emph{dataset diversity} over exhaustive kernel hyperparameter sweeps.
    \item Claims of quantum advantage based on a single or few datasets are inherently fragile, as our Phase~4 results demonstrate: the direction of the quantum--classical gap is consistent across seeds, but the dataset determines whether it favours quantum or classical.
\end{enumerate}

The non-linearity gap (RBF $-$ linear BA) shows a Spearman correlation with quantum advantage ($\rho = 0.683$, $p = 0.042$), suggesting potential for pre-screening: datasets where classical non-linearity is minimal may be better candidates for quantum kernels, aligning with the theoretical insight that quantum advantage should occur for problems that are hard for classical feature maps~\citep{liu2021rigorous, huang2021power}. We note that this correlation is based on only nine datasets and is at the boundary of statistical significance; it should be interpreted as a suggestive hypothesis rather than a confirmatory result.

\begin{figure}[t]
    \centering
    \includegraphics[width=0.7\columnwidth]{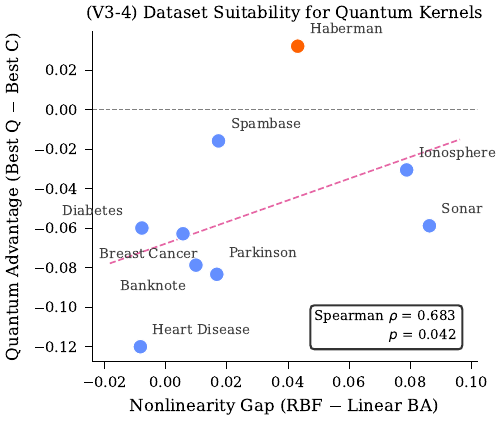}
    \caption{Dataset suitability criterion: non-linearity gap (RBF $-$ linear BA) vs.\ quantum advantage ($\Delta$). Spearman $\rho = 0.683$ ($p = 0.042$). Haberman is the only dataset above the dashed parity line, indicating positive quantum advantage.}
    \label{fig:suitability}
\end{figure}

\subsection{Hardware Validation and the Simulation--Reality Gap}
\label{sec:discussion_hardware}

The high kernel fidelity established in Sect.~\ref{sec:results_hardware} ($r \geq 0.976$ across all experiments) provides strong evidence that the negative results from our simulation-based benchmark transfer to real hardware at the tested circuit scales (3--5 qubits, depth 22--74).

The noise-as-regulariser trend observed in Sect.~\ref{sec:results_hardware} (4/6 experiments, mean $+1.5$~pp) aligns with theoretical work by \citet{heyraud2022noisy}, who showed that decoherence suppresses the effective kernel rank, acting as implicit regularisation. In classical machine learning, \citet{bishop1995training} established the formal equivalence between training with noise and Tikhonov regularisation; an analogous mechanism may operate at the kernel-computation level in quantum systems. However, the effect is small, not statistically significant at $n = 60$, and may not persist at larger sample sizes where regularisation is less critical.

\paragraph{Practical QPU readiness}
The experimental pipeline demonstrates end-to-end hardware readiness: vectorised SamplerV2 Primitive Unified Blocs (PUBs) enable efficient circuit submission ($\sim$1\,830 circuits per experiment) and per-batch checkpointing ensures resume-safety. No readout error mitigation is applied, yet the high kernel fidelity ($r \geq 0.976$) confirms that raw hardware counts suffice at the tested circuit scales. Each experiment consumes approximately 10 minutes of QPU time. Scaling to larger datasets ($n > 100$) will require $O(n^2)$ circuits, making QPU cost a primary bottleneck.

\subsection{Quantum Kernel Training: Promise and Limitations}
\label{sec:discussion_qkt}

QKT achieves the single competitive quantum result ($\BA = 0.968$ on breast\_cancer via belis/tree/k=8), almost matching the classical ceiling ($\BA = 0.976$). However, several concerns temper this success:

\begin{enumerate}
    \item \textbf{Computational cost}: QKT requires 50--170 iterations per fold, each involving a full kernel recomputation, resulting in $\sim2{,}000\times$ overhead vs.\ classical SVM.
    \item \textbf{Low convergence}: Only 13.6\% of QKT configurations converge (reach optimal KTA within tolerance). Most hit the iteration limit.
    \item \textbf{Theta instability}: Mean coefficient of variation of QKT parameters across folds is 0.54, indicating that the optimiser finds different solutions per fold. While KTA consistently improves (mean $+$0.13), the \emph{structure} of the learned scaling varies.
    \item \textbf{Dataset sensitivity}: QKT improves belis on 5/7 datasets but harms performance on small datasets (sonar: $-4$ to $-6$~pp regression), suggesting overfitting of the theta parameters.
    \item \textbf{Haberman failure}: QKT on haberman -- the sole dataset with quantum favourability -- yields $\BA = 0.500$--$0.586$ (near random), indicating that QKT cannot exploit the same structure that fixed quantum kernels capture.
\end{enumerate}

These findings suggest that QKT is best understood as a kernel bandwidth tuning method rather than a fundamental quantum advantage mechanism.

\subsection{Limitations}
\label{sec:discussion_limitations}

\paragraph{Simulation fidelity}
While our hardware experiments validate simulation at 3--5 qubits, the noise landscape at higher qubit counts may differ qualitatively. Coherent errors, crosstalk, and qubit leakage (i.e., population transfer to non-computational states) -- not captured by our depolarising noise model -- may become significant at $n > 10$.

\paragraph{Statistical power}
The Wilcoxon signed-rank test with $n = 5$ folds has a minimum achievable $p$-value of 0.0625, limiting power for small effect sizes. Our Phase~4 seed analysis partially addresses this, but the ideal approach would be repeated $k$-fold CV with more folds~or bootstrap resampling.

\paragraph{Preprocessing cost}
Our per-fold preprocessing design -- fitting scalers and dimensionality reducers independently on each outer training fold -- requires computing a separate kernel matrix for every outer fold rather than caching a single global $N \times N$ matrix. For quantum kernels this multiplies the already-dominant circuit budget by the number of outer folds. We accept this cost to ensure unbiased absolute performance estimates; future work on efficient per-fold kernel caching or feature-map parameter sharing may reduce this overhead.

\paragraph{Binary classification only}
All experiments address binary classification. The findings may not generalise to multi-class problems, regression, or structured prediction tasks.

\paragraph{Dataset scope}
Our nine datasets, while diverse, are drawn from the UCI repository and may not represent all relevant application domains. Industrial datasets with specific geometric structure may exhibit different quantum--classical dynamics.

\paragraph{Feature map selection}
We test four quantum feature maps, all based on single- and two-qubit gates with $\text{reps} \leq 3$. More expressive architectures (e.g., hardware-efficient circuit designs, problem-specific layouts) might perform differently.

\section{Conclusion}
\label{sec:conclusion}

We have presented a comprehensive, four-phase empirical study of quantum kernel support vector machines, benchmarking four quantum feature maps against three classical SVM kernels across nine binary classification datasets, with 970 total experiments and strict nested cross-validation (main benchmark: $5 \times 3$; extended study: $5 \times 5$).

\paragraph{Main findings}
Our results lead to the following principal conclusions:

\begin{enumerate}
    \item \textbf{No quantum advantage on standard tabular data.} None of the 29 pairwise comparisons reach statistical significance at $\alpha = 0.05$. The Phase~4 seed analysis (16 seeds, $n = 8{,}400$ SVM fits) confirms that classical kernels are significantly better on 6/7 datasets ($p < 0.001$). Only haberman shows robust quantum favourability ($p = 0.004$, 87.5\% seed win rate), with a modest $+1.9$~pp advantage.
    
    \item \textbf{Dataset choice dominates performance.} Kruskal--Wallis factorial analysis reveals that dataset selection accounts for 73\% of performance variance ($\varepsilon^2 = 0.73$), while kernel type contributes only 9\%. This implies that quantum--classical comparisons on a single dataset are insufficient.
    
    \item \textbf{Spectral mismatch explains the gap.} Quantum feature maps produce kernel matrices with extreme eigenspectra -- either near-identity (belis) or near-rank-1 (rot2dof) -- while the best classical kernel (RBF) achieves an intermediate profile. This ``Goldilocks hypothesis'' provides a constructive design criterion for future quantum feature maps.
    
    \item \textbf{Hardware validation confirms simulation faithfulness.} Across six experiments on IBM ibm\_fez (Heron~r2), kernel fidelity satisfies $r \geq 0.976$ (mean $r = 0.990$), and downstream SVM performance is statistically indistinguishable from simulation. The negative results from simulation transfer to hardware.
    
    \item \textbf{QKT yields the sole competitive result, at high cost.} Quantum kernel training via KTA optimisation reaches $\BA = 0.968$ on breast\_cancer, near the classical ceiling of $0.976$ -- but with $\sim2{,}000\times$ computational overhead, 13.6\% convergence rate, and high parameter instability.
    
    \item \textbf{Quantum kernels show steeper individual slopes but no aggregate data advantage.} Learning curves show steeper quantum slopes on 6/8 datasets and gap narrowing on 6/8; however, the mean quantum and classical slopes are equal (0.032), indicating that while quantum kernels start from a lower baseline they learn at a comparable aggregate rate.
\end{enumerate}

\paragraph{Recommendations}
Based on these findings, we offer the following recommendations for the quantum kernel research community:

\begin{itemize}
    \item \emph{For benchmarking}: Use nested cross-validation, multiple diverse datasets, and statistical significance testing with effect sizes. Report confidence intervals, not just point estimates.
    \item \emph{For feature map design}: Target intermediate spectral profiles. Compute the effective rank ratio as a cheap diagnostic before committing to expensive evaluations.
    \item \emph{For QKT}: Invest in more efficient optimisation strategies (e.g., analytic gradients, quantum natural gradient) to reduce the $\sim2{,}000\times$ overhead. Address convergence instability.
    \item \emph{For advantage claims}: Demonstrate improvements that are (i)~statistically significant, (ii)~substantial (${>}5$~pp), (iii)~consistent across seeds, and (iv)~observed on multiple datasets.
\end{itemize}

\paragraph{Future work}
Several directions merit further investigation:
\begin{itemize}
    \item \emph{Hardware scaling}: Extending hardware validation to higher qubit counts ($n = 10{-}20$) and more complex feature maps to test whether fidelity degrades gracefully or catastrophically.
    \item \emph{Problem-specific feature maps}: Designing quantum circuits informed by the data geometry or symmetry group~\citep{glick2024covariant}, rather than using generic architectures.
    \item \emph{Projected quantum kernels}: Using partial trace operations to create richer kernel structures that may achieve better spectral properties~\citep{huang2021power}.
    \item \emph{Error mitigation}: Evaluating whether advanced error mitigation techniques (zero-noise extrapolation, probabilistic error cancellation) can improve hardware kernel quality beyond readout correction.
    \item \emph{Real-world industrial datasets}: Testing quantum kernels on domain-specific classification problems (e.g.~manufacturing quality control, predictive maintenance) where data geometry may differ from standard UCI benchmarks.
\end{itemize}

\paragraph{Reproducibility}
The complete benchmark suite -- comprising $\sim$18\,700 lines of open-source Python code, 683 cached kernel matrices, 54 figures, and 132 unit tests -- is publicly available to facilitate reproduction and extension of our results.


\appendix

\section{Dataset Descriptions}
\label{app:datasets}

Table~\ref{tab:datasets_full} provides detailed descriptions of all nine binary classification datasets used in this study. All datasets are publicly available from the UCI Machine Learning Repository~\citep{uci2017} or OpenML~\citep{vanschoren2014openml}.

\begin{table}[htbp]
\centering
\caption{Full dataset descriptions, preprocessing, and class distributions. $N$: post-cleaning sample count used in all experiments; $d$: original feature dimensionality.}
\label{tab:datasets_full}
\small
\begin{tabularx}{\textwidth}{@{}l r r l X@{}}
\toprule
Dataset & $N$ & $d$ & Class ratio & Description \\
\midrule
banknote        & \num{1372} & 4  & 56:44 & Wavelet-transformed features from banknote images; task is genuine vs.\ forged classification. \\[3pt]
breast\_cancer  & 569    & 30 & 63:37 & Wisconsin Diagnostic Breast Cancer; 30 real-valued features computed from digitised cell nuclei images. \\[3pt]
diabetes\_pima  & 768    & 8  & 65:35 & Diabetes diagnosis from clinical measurements (Pima community); 8 features. \\[3pt]
haberman        & 306    & 3  & 74:26 & Haberman's survival data; 3 features: patient age, year of operation, number of axillary nodes. \\[3pt]
heart\_disease  & 303    & 13 & 54:46 & Cleveland Heart Disease; 13 clinical attributes including chest pain type, resting blood pressure. \\[3pt]
ionosphere      & 351    & 34 & 64:36 & Johns Hopkins Ionosphere radar returns; 34 pulse features indicating good/bad radar signal. \\[3pt]
parkinson\_489  & 489    & 22 & 75:25 & Replicated Acoustic Features -- Parkinson Database; 22 acoustic features from sustained phonation tasks (80~subjects, grouped by speaker for CV). \\[3pt]
sonar           & 208    & 60 & 53:47 & Connectionist Bench; 60 frequency-band energy features from sonar returns (rocks vs.\ mines). \\[3pt]
spambase        & \num{4601} & 57 & 61:39 & UCI Spambase; 57 word/character frequency features for spam vs.\ ham email classification. \\
\botrule
\end{tabularx}
\end{table}

\subsection*{Preprocessing Pipeline}

All datasets undergo the following standardised preprocessing pipeline, applied \emph{independently per outer CV fold} to prevent information leakage:

\begin{enumerate}
    \item \textbf{Missing value handling}: Datasets with missing values (diabetes\_pima: zeros as missing for glucose, blood pressure, body mass index (BMI)) are imputed with column medians.
    \item \textbf{Standardisation}: Features are scaled to zero mean and unit variance using \texttt{StandardScaler} (or \texttt{MinMaxScaler} for NMF), fit on the outer-fold training split only and then applied to both training and test splits.
    \item \textbf{Dimensionality reduction}: Features are projected to $k \in \{3, 4, 6, 8, 10, 12, 14\}$ components using one of three methods, each fit on the training split:
    \begin{itemize}
        \item \textbf{PCA}: Principal Component Analysis, retaining the top-$k$ principal components.
        \item \textbf{NMF}: Non-negative Matrix Factorisation~\citep{lee1999learning}, extracting $k$ non-negative components.
        \item \textbf{Tree}: \texttt{DecisionTreeClassifier}-based feature importance ranking, selecting the top-$k$ features by Gini impurity.
    \end{itemize}
    \item \textbf{Kernel computation}: Reduced features are passed directly as rotation-gate angles to the quantum feature maps (no further rescaling to a fixed range is applied). Kernel matrices are computed per fold from the transformed features.
\end{enumerate}

\subsection*{Subsampling for Hardware Experiments}

For Phase~3 hardware validation (Sect.~\ref{sec:results_hardware}), full datasets are subsampled to $n = 60$ stratified samples to fit within QPU time budgets. Subsampling preserves the original class ratio via stratified random sampling with fixed seed ($\texttt{seed} = 42$). The subsampled indices are stored in \texttt{split\_indices.json} for reproducibility.

\subsection*{Dataset Provenance}

Table~\ref{tab:dataset_provenance} lists the primary repository identifiers for each dataset. The benchmark framework fetches datasets at runtime using the \texttt{ucimlrepo} Python package (primary source) with an automatic fallback to OpenML~\citep{vanschoren2014openml} via \texttt{sklearn.datasets.fetch\_openml} (specifying \texttt{version=1} for reproducibility). The breast\_cancer dataset is loaded from scikit-learn's built-in copy, which mirrors the UCI source. Downloaded datasets are cached locally by the respective libraries (\texttt{ucimlrepo} and scikit-learn) and are not re-fetched on subsequent runs.

\begin{table}[htbp]
\centering
\caption{Dataset provenance. UCI~ID: identifier used by \texttt{ucimlrepo.fetch\_ucirepo()}; OpenML name: identifier used by \texttt{sklearn.datasets.fetch\_openml()} as fallback.}
\label{tab:dataset_provenance}
\small
\begin{tabular}{@{}llrl@{}}
\toprule
\textbf{Dataset} & \textbf{Primary source} & \textbf{UCI~ID} & \textbf{OpenML name} \\
\midrule
banknote        & UCI  & 267 & banknote-authentication \\
breast\_cancer  & sklearn built-in & --- & --- \\
diabetes\_pima  & UCI  &  34 & diabetes \\
haberman        & UCI  &  43 & haberman \\
heart\_disease  & UCI  &  45 & heart-statlog \\
ionosphere      & UCI  &  52 & ionosphere \\
parkinson\_489  & UCI  & 489 & (local CSV fallback) \\
sonar           & UCI  & 151 & sonar \\
spambase        & UCI  &  94 & spambase \\
\botrule
\end{tabular}
\end{table}

\section{Quantum Feature Map Specifications}
\label{app:feature_maps}

This appendix details the four quantum feature maps evaluated in this study. All circuits use the compute-uncompute (fidelity) kernel: $K(\bm{x}, \bm{x}') = |\braket{\psi(\bm{x})}{\psi(\bm{x}')}|^2$, where $\ket{\psi(\bm{x})} = U(\bm{x})\ket{0}^{\otimes n}$.

\textbf{Encoding convention.} Rot2DoF, Belis, and Sakhnenko10 employ \emph{double-feature encoding}: each qubit absorbs two classical features via paired rotations, so $n = \lceil k/2 \rceil$ qubits suffice for $k$ features. ZZFeatureMap uses \emph{single-feature encoding} with $n = k$ qubits. Property tables below list $k$ as the feature count; see Section~\ref{sec:methodology_preprocessing} for qubit counts.

\begin{figure}[htbp]
    \centering
    \includegraphics[width=\textwidth]{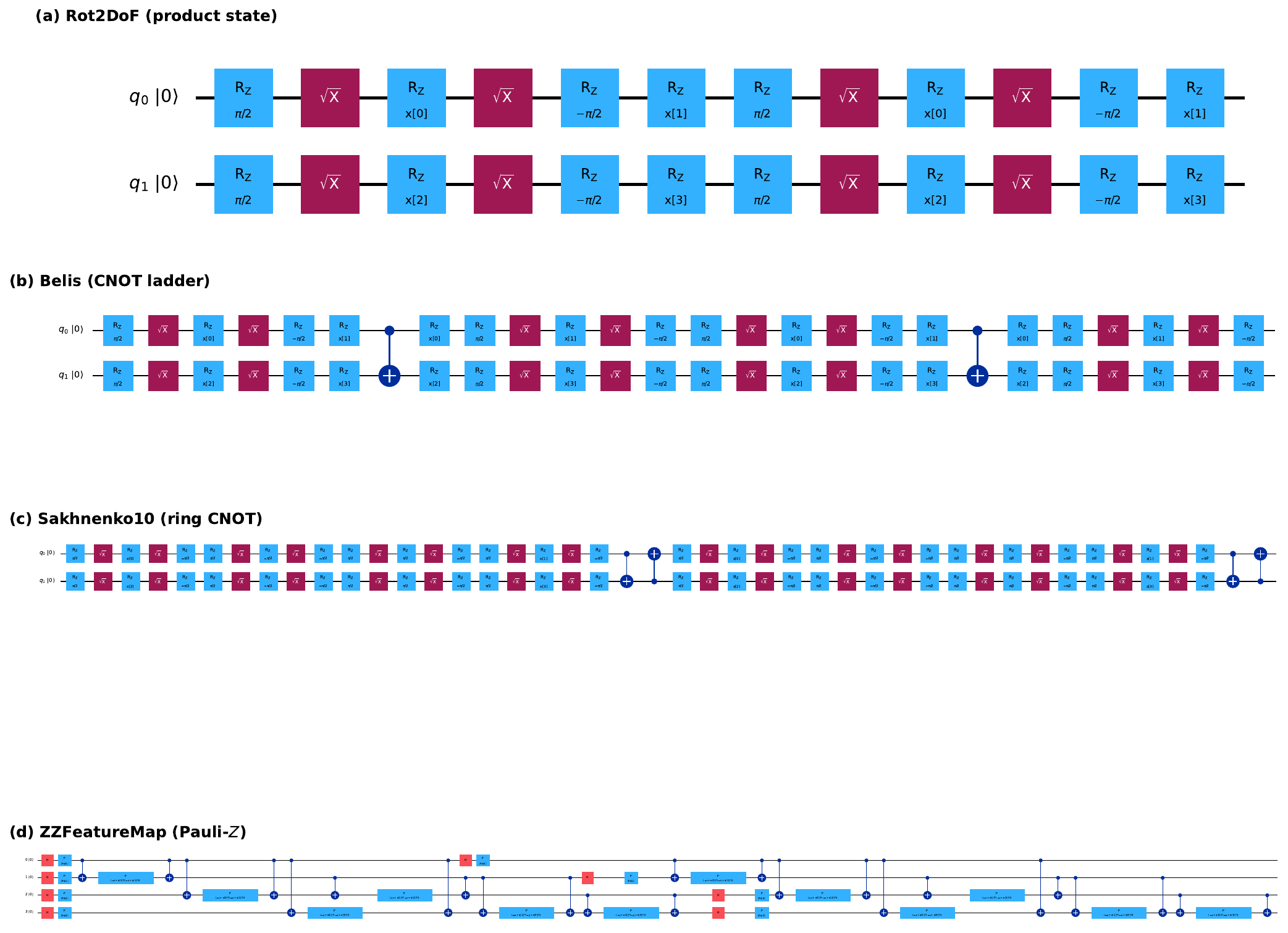}
    \caption{Quantum circuits for the four evaluated feature maps, shown at $k = 4$ features. (a)~Rot2DoF uses product-state encoding with $R_X$--$R_Z$ rotations and no entanglement ($n = 2$ qubits). (b)~Belis applies $R_X$--$R_Z$ encoding with a CNOT ladder for linear-connectivity entanglement ($n = 2$ qubits). (c)~Sakhnenko10 uses ring-topology CNOT gates with $R_X$ data encoding ($n = 2$ qubits). (d)~ZZFeatureMap encodes data through pairwise $ZZ$ interactions ($n = 4$ qubits). Rot2DoF/Belis/Sakhnenko10 require $\lceil k/2 \rceil$ qubits; ZZFeatureMap requires $k$ qubits.}
    \label{fig:circuits_all}
\end{figure}

Figure~\ref{fig:circuits_all} shows the quantum circuits for the four evaluated feature maps at $k = 4$ features ($n = 2$ qubits for rot2dof, belis, and sakhnenko10; $n = 4$ qubits for zzfm).

\subsection{Rot2DoF (Two-Parameter Rotation)}
\label{app:rot2dof}

A product-state encoding that applies two single-qubit rotations per qubit per repetition:
\begin{equation}
    U_{\text{rot2dof}}(\bm{x}) = \prod_{r=1}^{R} \prod_{q=0}^{n-1} R_X(x_{2q})\, R_Z(x_{2q+1})
\end{equation}
where $R_X$ and $R_Z$ are Pauli rotation gates, $R$ is the number of repetitions, and qubit $q$ encodes features $x_{2q}$ and $x_{2q+1}$ (double-feature encoding). Despite its simplicity, rot2dof achieves competitive QSVM performance due to its expressiveness in the single-qubit Bloch sphere.

\begin{table}[htbp]
\centering
\caption{Rot2DoF circuit properties at $R=2$ repetitions.}
\small
\begin{tabular}{@{}lcccc@{}}
\toprule
$k$ (features) & Depth & CX/CZ gates & Gate slots & Entanglement \\
\midrule
4  & 12 & 0 & 16 & None \\
8  & 12 & 0 & 32 & None \\
14 & 12 & 0 & 56 & None \\
\botrule
\end{tabular}
\end{table}

\subsection{Belis (CNOT Ladder)}
\label{app:belis}

Proposed by \citet{belis2024quantum} for high-energy physics jet classification. Each repetition applies an encoding block~$S_x$ with $R_X(x_{2q})\, R_Z(x_{2q+1})$ per qubit, a CNOT ladder for nearest-neighbour entanglement, and a second encoding block~$S'_x$ with swapped rotation axes ($R_Z(x_{2q})\, R_X(x_{2q+1})$):
\begin{equation}
    U_{\text{belis}}(\bm{x}) = \prod_{r=1}^{R} \left[ S'_x \cdot \prod_{q=0}^{n-2} \text{CX}_{q,q+1} \cdot S_x \right]
\end{equation}
where $S_x = \prod_{q=0}^{n-1} R_X(x_{2q})\, R_Z(x_{2q+1})$ and $S'_x = \prod_{q=0}^{n-1} R_Z(x_{2q})\, R_X(x_{2q+1})$.

The CNOT ladder creates linear-connectivity entanglement, producing near-identity kernels with high effective rank (0.40--0.74) but limited discriminative power due to eigenspectral flatness.

\begin{table}[htbp]
\centering
\caption{Belis circuit properties at $R=2$ repetitions.}
\small
\begin{tabular}{@{}lcccc@{}}
\toprule
$k$ (features) & Depth & CX/CZ gates & Gate slots & Entanglement \\
\midrule
4  & 26 & 2  & 8  & Linear \\
8  & 29 & 6  & 16 & Linear \\
14 & 32 & 12 & 28 & Linear \\
\botrule
\end{tabular}
\end{table}

\subsection{Sakhnenko (10-Parameter Ansatz)}
\label{app:sakhnenko10}

Based on \citet{sakhnenko2022hybrid}, this feature map uses double-feature $R_X$ encoding with fixed intermediate rotations and ring entanglement:
\begin{equation}
    U_{\text{sakh.}}(\bm{x}) = \prod_{r=1}^{R} \left[ W_{\text{ent}} \cdot \prod_{q=0}^{n-1} R_X(x_{2q})\, R_Y(\tfrac{\pi}{2})\, R_X(\tfrac{\pi}{2})\, R_X(x_{2q+1}) \right]
\end{equation}
where $W_{\text{ent}}$ applies CX gates in a ring topology ($q_0 \to q_1 \to \cdots \to q_{n-1} \to q_0$).

\begin{table}[htbp]
\centering
\caption{Sakhnenko10 circuit properties at $R=2$ repetitions.}
\small
\begin{tabular}{@{}lcccc@{}}
\toprule
$k$ (features) & Depth & CX/CZ gates & Gate slots & Entanglement \\
\midrule
4  & 44 & 4  & 16 & Ring \\
8  & 48 & 8  & 32 & Ring \\
14 & 54 & 14 & 56 & Ring \\
\botrule
\end{tabular}
\end{table}

\subsection{ZZFeatureMap (Qiskit Standard)}
\label{app:zzfm}

The standard Qiskit feature map~\citep{javadi2024quantum} based on the approach of \citet{havlicek2019supervised}. Encodes data through $ZZ$ interactions:
\begin{equation}
    U_{\text{ZZ}}(\bm{x}) = \prod_{r=1}^{R} \left[ \prod_{i<j} \exp\!\left(i\, x_i x_j\, Z_i Z_j\right) \prod_{i=1}^{n} H_i \right]
\end{equation}
where $H$ denotes the Hadamard gate. The ZZ interaction terms create pairwise entanglement modulated by the product of feature values.

\begin{table}[htbp]
\centering
\caption{ZZFeatureMap circuit properties at $R=2$ repetitions.}
\small
\begin{tabular}{@{}lcccc@{}}
\toprule
$k$ (features) & Depth & CX/CZ gates & Gate slots & Entanglement \\
\midrule
4  & 31 & 24  & 4  & Full (via $ZZ$) \\
8  & 67 & 112 & 8  & Full (via $ZZ$) \\
14 & 121 & 364 & 14 & Full (via $ZZ$) \\
\botrule
\end{tabular}
\end{table}
The reported depth and CX/CZ gate counts reflect the parametrised and decomposed circuit properties with full entanglement before transpilation.

\subsection{Transpiled Circuit Properties on ibm\_fez}

For the hardware experiments (Phase~3), circuits were transpiled to the ibm\_fez native gate set $\{rz, sx, cz, x\}$ at \texttt{optimization\_level=3}. Table~\ref{tab:transpiled_circuits} shows the post-transpilation properties for the two feature maps used in hardware experiments.

\begin{table}[htbp]
\centering
\caption{Transpiled circuit properties on ibm\_fez (Heron~r2) for hardware experiments. Entries marked ``--'' were not separately profiled; only the qubit count and depth were verified post-transpilation.}
\label{tab:transpiled_circuits}
\small
\begin{tabular}{@{}llccccccc@{}}
\toprule
Feature Map & $k$ & Depth & CZ & RZ & SX & X & Measurements \\
\midrule
Belis   & 10 & 74 & 16 & 137 & 82 & 10 & 5 \\
Rot2DoF & 10 & 22 &  0 &  65 & 40 &  0 & 5 \\
Belis   &  6 & 62 & 10 & --  & -- & -- & 3 \\
Rot2DoF &  6 & 22 &  0 &  39 & 24 &  0 & 3 \\
\botrule
\end{tabular}
\end{table}

\noindent Note: rot2dof transpiles to zero CZ gates on Heron~r2, confirming that this feature map creates only product states despite the ``entangling rotations'' in its parametric form. The aggressive instruction-set-architecture (ISA) transpilation at level~3 eliminates redundant two-qubit gates.

\section{Extended Hardware Validation Results}
\label{app:hardware}
\setcounter{figure}{0}

This appendix provides per-experiment details for all six hardware experiments executed on IBM ibm\_fez (Heron~r2, 156 qubits). All experiments use 1\,024 shots per circuit without readout error mitigation; kernel entries are computed from raw counts. Fidelity tables report mean absolute error (MAE), root mean squared error (RMSE), and relative Frobenius norm between hardware and reference kernel matrices.

\subsection[HW-01: Breast Cancer / Belis / NMF / k=10]{HW-01: Breast Cancer / Belis / NMF / $k{=}10$}
\label{app:hw01}

\begin{table}[htbp]
\centering
\caption{HW-01 kernel fidelity and SVM performance.}
\small
\begin{tabular}{@{}lccccc@{}}
\toprule
Reference & Pearson $r$ & Spearman $\rho$ & MAE & RMSE & Frob.\ (rel.) \\
\midrule
Ideal (statevector) & 0.9858 & 0.9643 & 0.0075 & 0.0109 & 0.0736 \\
Noisy (depolarising) & 0.9860 & 0.9668 & 0.0066 & 0.0097 & 0.2723 \\
\botrule
\end{tabular}
\end{table}

\begin{table}[htbp]
\centering
\caption{HW-01 downstream SVM balanced accuracy (5-fold stratified CV, $n{=}60$).}
\small
\begin{tabular}{@{}lcr@{}}
\toprule
Kernel source & BA (mean $\pm$ std) & Best $C$ \\
\midrule
Hardware (ibm\_fez) & $0.473 \pm 0.033$ & 10.0 \\
Ideal (statevector) & $0.461 \pm 0.051$ & 10.0 \\
Noisy (depolarising) & $0.461 \pm 0.051$ & 10.0 \\
\botrule
\end{tabular}
\end{table}

\noindent\textbf{Kernel properties}: 60$\times$60, off-diagonal mean $0.045 \pm 0.049$, effective rank 54.25/60 (90.4\%). Near-identity kernel with high effective rank, consistent with belis spectral profile (Sect.~\ref{sec:discussion_spectral}).

\noindent\textbf{Execution}: 1\,830 circuits, 7 batches, 681~s wall time. Transpiled depth 74, 16 CZ gates.

\subsection[HW-02: Breast Cancer / Rot2DoF / NMF / k=10]{HW-02: Breast Cancer / Rot2DoF / NMF / $k{=}10$}
\label{app:hw02}

\begin{table}[htbp]
\centering
\caption{HW-02 kernel fidelity and SVM performance.}
\small
\begin{tabular}{@{}lccccc@{}}
\toprule
Reference & Pearson $r$ & Spearman $\rho$ & MAE & RMSE & Frob.\ (rel.) \\
\midrule
Ideal (statevector) & 0.9961 & 0.9955 & 0.0176 & 0.0223 & --- \\
Noisy (depolarising) & 0.9961 & 0.9956 & 0.0238 & 0.0302 & --- \\
\botrule
\end{tabular}
\end{table}

\begin{table}[htbp]
\centering
\caption{HW-02 downstream SVM balanced accuracy (5-fold stratified CV, $n{=}60$).}
\small
\begin{tabular}{@{}lcr@{}}
\toprule
Kernel source & BA (mean $\pm$ std) & Best $C$ \\
\midrule
Hardware (ibm\_fez) & $0.760 \pm 0.131$ & 10.0 \\
Ideal (statevector) & $0.686 \pm 0.086$ & 100.0 \\
Noisy (depolarising) & $0.686 \pm 0.086$ & 100.0 \\
\botrule
\end{tabular}
\end{table}

\noindent\textbf{Key observation}: $\Delta(\text{HW} - \text{Ideal}) = +7.46$~pp. Hardware noise acts as a regulariser for rot2dof, selecting a lower $C$ (10 vs.\ 100) and achieving higher BA. Transpiled depth 22, zero CZ gates on Heron~r2.

\subsection[HW-03: Sonar / Belis / PCA / k=6]{HW-03: Sonar / Belis / PCA / $k{=}6$}
\label{app:hw03}

\begin{table}[htbp]
\centering
\caption{HW-03 kernel fidelity and SVM performance.}
\small
\begin{tabular}{@{}lccccc@{}}
\toprule
Reference & Pearson $r$ & Spearman $\rho$ & MAE & RMSE \\
\midrule
Ideal (statevector) & 0.9843 & 0.9795 & 0.0179 & 0.0233 \\
Noisy (depolarising) & 0.9847 & 0.9802 & 0.0146 & 0.0205 \\
\botrule
\end{tabular}
\end{table}

\begin{table}[htbp]
\centering
\caption{HW-03 downstream SVM balanced accuracy (5-fold stratified CV, $n{=}60$).}
\small
\begin{tabular}{@{}lcr@{}}
\toprule
Kernel source & BA (mean $\pm$ std) & Best $C$ \\
\midrule
Hardware (ibm\_fez) & $0.550 \pm 0.114$ & 10.0 \\
Ideal (statevector) & $0.516 \pm 0.063$ & 10.0 \\
Noisy (depolarising) & $0.520 \pm 0.032$ & 1.0 \\
\botrule
\end{tabular}
\end{table}

\noindent\textbf{Cross-dataset validation}: Belis fidelity on sonar ($r = 0.984$) is comparable to breast\_cancer ($r = 0.986$), confirming cross-dataset generalisation.

\subsection[HW-04: Sonar / Rot2DoF / PCA / k=6]{HW-04: Sonar / Rot2DoF / PCA / $k{=}6$}
\label{app:hw04}

\begin{table}[htbp]
\centering
\caption{HW-04 kernel fidelity (hardware vs.\ reference).}
\small
\begin{tabular}{@{}lcccc@{}}
\toprule
Reference & Pearson $r$ & MAE & RMSE & Frob.\ rel. \\
\midrule
Ideal (statevector) & 0.9980 & 0.0126 & 0.0168 & 0.041 \\
\botrule
\end{tabular}
\end{table}

\begin{table}[htbp]
\centering
\caption{HW-04 downstream SVM balanced accuracy (5-fold stratified CV, $n{=}60$).}
\small
\begin{tabular}{@{}lcr@{}}
\toprule
Kernel source & BA (mean $\pm$ std) & Best $C$ \\
\midrule
Hardware (ibm\_fez) & $0.544 \pm 0.095$ & 0.01 \\
Ideal (statevector) & $0.570 \pm 0.122$ & 100.0 \\
Noisy (depolarising) & $0.570 \pm 0.122$ & 1.0 \\
\botrule
\end{tabular}
\end{table}

\noindent\textbf{Entanglement comparison (sonar)}: Rot2dof ($r = 0.998$) vs.\ belis ($r = 0.984$) on sonar—the shallower rot2dof circuit achieves higher fidelity, consistent with breast\_cancer results.

\subsection[HW-05: Diabetes Pima / Belis / NMF / k=6]{HW-05: Diabetes Pima / Belis / NMF / $k{=}6$}
\label{app:hw05}

\begin{table}[htbp]
\centering
\caption{HW-05 kernel fidelity (hardware vs.\ reference).}
\small
\begin{tabular}{@{}lcccc@{}}
\toprule
Reference & Pearson $r$ & MAE & RMSE & Frob.\ rel. \\
\midrule
Ideal (statevector) & 0.9760 & 0.0250 & 0.0340 & 0.136 \\
\botrule
\end{tabular}
\end{table}

\begin{table}[htbp]
\centering
\caption{HW-05 downstream SVM balanced accuracy (5-fold stratified CV).}
\small
\begin{tabular}{@{}lcr@{}}
\toprule
Kernel source & BA (mean $\pm$ std) & Best $C$ \\
\midrule
Hardware (ibm\_fez) & $0.511 \pm 0.091$ & 10.0 \\
Ideal (statevector) & $0.566 \pm 0.115$ & 10.0 \\
Noisy (depolarising) & $0.545 \pm 0.095$ & 10.0 \\
\botrule
\end{tabular}
\end{table}

\noindent\textbf{Third dataset validated}: Belis fidelity on diabetes\_pima ($r = 0.976$) is the lowest of the three datasets (breast\_cancer: 0.986, sonar: 0.984), but still well above the $r = 0.95$ threshold typically considered adequate for downstream task transfer. The higher MAE (0.025) reflects the more complex feature geometry of the diabetes dataset.

\subsection[HW-06: Diabetes Pima / Rot2DoF / NMF / k=6]{HW-06: Diabetes Pima / Rot2DoF / NMF / $k{=}6$}
\label{app:hw06}

\begin{table}[htbp]
\centering
\caption{HW-06 kernel fidelity (hardware vs.\ reference).}
\small
\begin{tabular}{@{}lcccc@{}}
\toprule
Reference & Pearson $r$ & MAE & RMSE & Frob.\ rel. \\
\midrule
Ideal (statevector) & 0.9970 & 0.0170 & 0.0220 & 0.032 \\
\botrule
\end{tabular}
\end{table}

\begin{table}[htbp]
\centering
\caption{HW-06 downstream SVM balanced accuracy (5-fold stratified CV).}
\small
\begin{tabular}{@{}lcr@{}}
\toprule
Kernel source & BA (mean $\pm$ std) & Best $C$ \\
\midrule
Hardware (ibm\_fez) & $0.538 \pm 0.075$ & 0.01 \\
Ideal (statevector) & $0.486 \pm 0.029$ & 10.0 \\
Noisy (depolarising) & $0.531 \pm 0.067$ & 10.0 \\
\botrule
\end{tabular}
\end{table}

\noindent\textbf{Noise-as-regulariser observation}: $\Delta = +5.2$~pp, the second largest hardware improvement after HW-02 (+7.5~pp). Both cases involve the rot2dof feature map, suggesting that product-state encodings may benefit more from hardware noise regularisation, though this effect is not statistically significant at $n = 60$.

\subsection{Cross-Experiment Summary}

\begin{table}[htbp]
\centering
\caption{Summary of all six hardware experiments on ibm\_fez.}
\label{tab:hw_summary_full}
\small
\begin{tabular}{@{}llllcrrrr@{}}
\toprule
ID & Dataset & FM & $k$ & Depth & $r$ & MAE & BA\textsubscript{HW} & $\Delta$ \\
\midrule
HW-01 & breast\_cancer & belis   & 10 & 74 & 0.986 & 0.008 & 0.473 & $+1.2$ \\
HW-02 & breast\_cancer & rot2dof & 10 & 22 & 0.996 & 0.018 & 0.760 & $+7.5$ \\
HW-03 & sonar          & belis   &  6 & 62 & 0.984 & 0.018 & 0.550 & $+3.3$ \\
HW-04 & sonar          & rot2dof &  6 & 22 & 0.998 & 0.013 & 0.544 & $-2.6$ \\
HW-05 & diabetes\_pima & belis   &  6 & 62 & 0.976 & 0.025 & 0.511 & $-5.5$ \\
HW-06 & diabetes\_pima & rot2dof &  6 & 22 & 0.997 & 0.017 & 0.538 & $+5.2$ \\
\botrule
\end{tabular}
\end{table}

\noindent\textbf{Aggregate statistics}: Mean $r = 0.990$ (min 0.976, max 0.998). Mean MAE $= 0.016$. Mean $|\Delta| = 4.2$~pp (mean signed $\Delta = +1.5$~pp). Rot2dof achieves higher fidelity on all three datasets (mean $r = 0.997$ vs.\ 0.982 for belis), attributable to its shallower transpiled circuits.

\noindent\textbf{Research questions answered}:
\begin{itemize}
    \item \textbf{Q1} (Headline fidelity): $r \geq 0.976$ across all experiments. Simulation is a faithful proxy for hardware.
    \item \textbf{Q2} (Entanglement effect): Dataset-dependent. Rot2dof outperforms belis by +28.7~pp on breast\_cancer, but only +2.6~pp on diabetes\_pima and $-0.6$~pp on sonar. Rot2dof consistently achieves higher kernel fidelity ($r \geq 0.997$).
    \item \textbf{Q3} (Cross-dataset): Both feature maps maintain high fidelity across all three datasets. Belis: $r \in [0.976, 0.986]$. Rot2dof: $r \in [0.996, 0.998]$.
\end{itemize}

\begin{figure}[htbp]
    \centering
    \includegraphics[width=\textwidth]{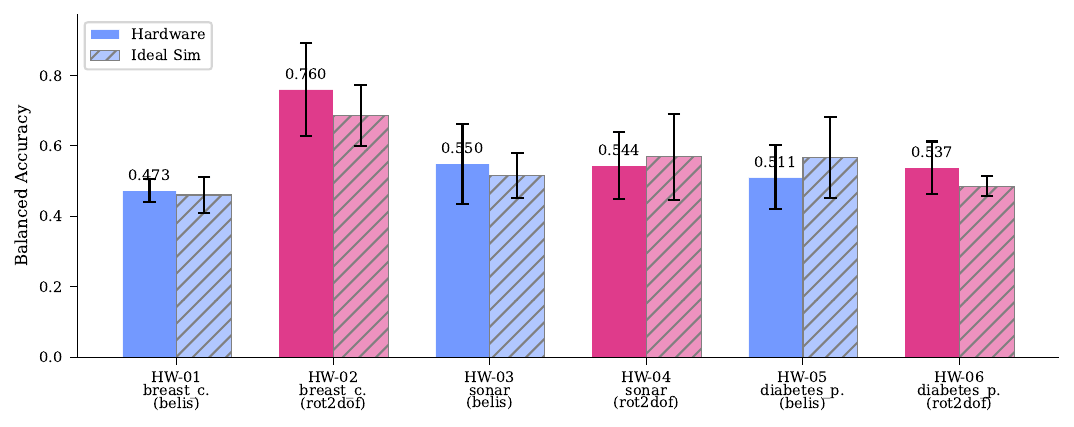}
    \caption{SVM balanced accuracy across all six hardware experiments.
        Solid bars: hardware (ibm\_fez, 1\,024 shots);
        hatched bars: ideal statevector simulation.
        Colour encodes the feature map; x-axis labels identify the
        experiment, dataset, and feature map.
        Rot2dof achieves the highest BA on breast\_cancer (HW-02),
        while hardware noise acts as a regulariser on four of six
        experiments ($\Delta > 0$).}
    \label{fig:hw_svm_performance}
\end{figure}

\begin{figure}[htbp]
    \centering
    \includegraphics[width=\textwidth]{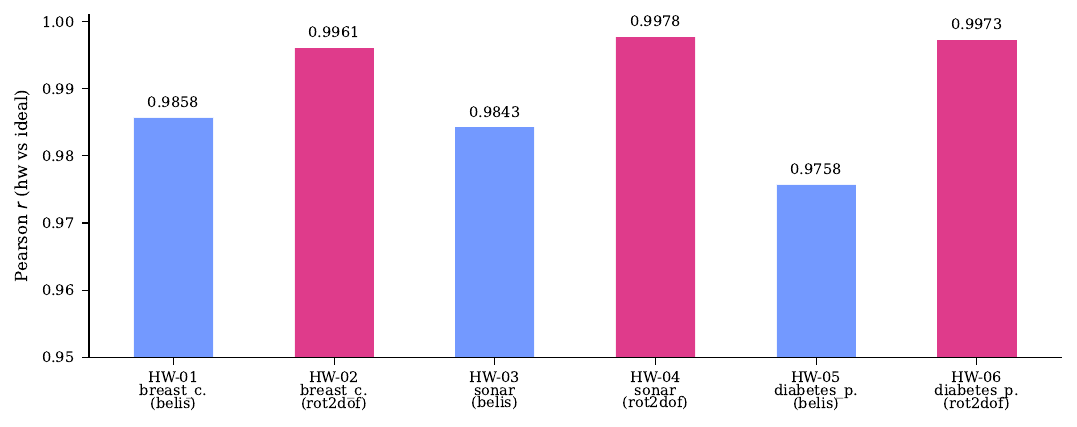}
    \caption{Kernel fidelity (Pearson~$r$) of the hardware kernel matrix
        versus ideal statevector simulation for each experiment.
        All experiments exceed $r = 0.97$.
        Rot2dof (product-state circuits) consistently achieves higher
        fidelity than belis (entangling circuits), reflecting its
        shallower transpiled depth on Heron~r2.}
    \label{fig:hw_kernel_fidelity}
\end{figure}

\begin{figure}[htbp]
    \centering
    \includegraphics[width=\textwidth]{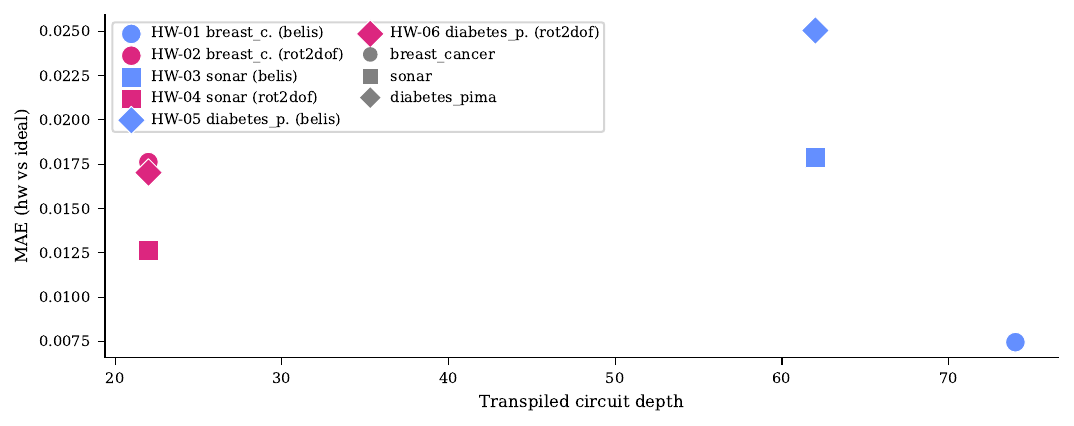}
    \caption{Kernel mean absolute error (MAE) versus transpiled circuit
        depth for all six hardware experiments.
        Colour encodes the feature map (blue: belis, pink: rot2dof);
        marker shape encodes the dataset (circle: breast\_cancer,
        square: sonar, diamond: diabetes\_pima).
        Deeper circuits (belis, depth $\geq 62$) incur higher MAE
        than shallow circuits (rot2dof, depth~22), confirming that
        circuit depth is the primary driver of hardware noise.}
    \label{fig:hw_noise_vs_depth}
\end{figure}

\begin{figure}[htbp]
    \centering
    \includegraphics[width=\textwidth]{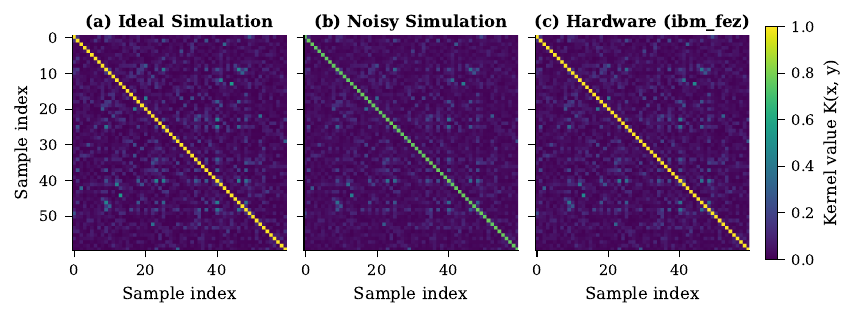}
    \caption{Kernel matrix heatmaps for HW-01
        (belis, breast\_cancer, $k{=}10$).
        Left to right: ideal statevector simulation, depolarising
        noise simulation, and IBM ibm\_fez hardware.
        The belis kernel is near-identity with off-diagonal values
        $\approx 0.045$; the hardware reproduces this structure
        with Pearson~$r = 0.986$.}
    \label{fig:hw_kernel_heatmap_hw01}
\end{figure}

\begin{figure}[htbp]
    \centering
    \includegraphics[width=\textwidth]{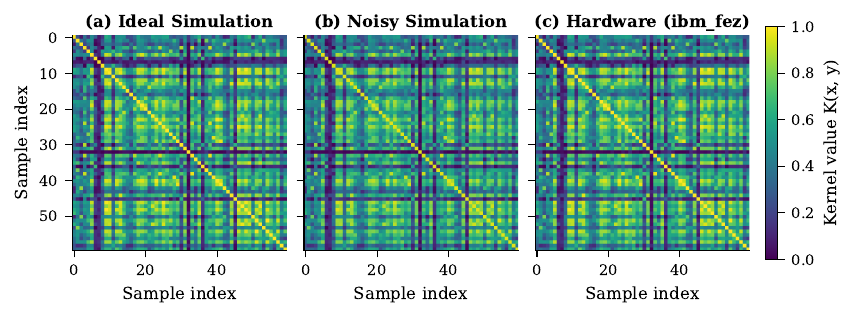}
    \caption{Kernel matrix heatmaps for HW-02
        (rot2dof, breast\_cancer, $k{=}10$).
        Left to right: ideal statevector simulation, depolarising
        noise simulation, and IBM ibm\_fez hardware.
        Rot2dof produces richer off-diagonal structure than belis,
        and the hardware achieves Pearson~$r = 0.996$, the highest
        fidelity among all experiments.}
    \label{fig:hw_kernel_heatmap_hw02}
\end{figure}

\section{Full Result Tables}
\label{app:tables}

This appendix presents the complete result tables referenced in the main text, including per-dataset detailed results, cost breakdowns, and learning curve data.

\subsection{Pairwise Wilcoxon Signed-Rank Tests}
\label{app:wilcoxon}

Table~\ref{tab:wilcoxon_full} reports the 9 best-vs-best pairwise Wilcoxon signed-rank tests comparing quantum and classical kernel configurations on balanced accuracy across 5~outer CV folds. The full set of 29 comparisons (multiple quantum--classical pairs) is consistent with the results shown here: none reach significance at $\alpha = 0.05$.

\begin{table}[htbp]
\centering
\caption{Pairwise Wilcoxon signed-rank tests: best quantum ideal vs.\ best classical, per dataset. All tests use 5~paired observations (outer folds). This table shows the 9 best-vs-best comparisons; the full set of 29 pairwise comparisons (multiple quantum--classical pairs per dataset) is available in the supplementary data.}
\label{tab:wilcoxon_full}
\small
\begin{tabular}{@{}llllrrl@{}}
\toprule
Dataset & Best Quantum & BA\textsubscript{Q} & Best Classical & BA\textsubscript{C} & $p$-value & Sig.? \\
\midrule
haberman       & sakhnenko10 & 0.581 & rbf     & 0.549 & 0.3125 & No \\
spambase       & belis       & 0.888 & rbf     & 0.903 & 0.0625 & No \\
ionosphere     & belis       & 0.905 & rbf     & 0.935 & 0.0625 & No \\
sonar          & rot2dof     & 0.813 & rbf     & 0.872 & 0.1250 & No \\
diabetes\_pima & rot2dof     & 0.669 & linear  & 0.729 & 0.0625 & No \\
breast\_cancer & rot2dof     & 0.913 & rbf     & 0.976 & 0.0625 & No \\
banknote       & rot2dof     & 0.921 & rbf     & 1.000 & 0.0625 & No \\
parkinson\_489 & rot2dof     & 0.746 & rbf     & 0.829 & 0.1250 & No \\
heart\_disease & rot2dof     & 0.724 & poly3   & 0.844 & 0.0625 & No \\
\botrule
\end{tabular}
\end{table}

\noindent Note: The minimum achievable two-sided $p$-value under the exact Wilcoxon signed-rank test with $n = 5$ paired observations is $p = 0.0625$, attained when all five fold differences share the same sign (rank sum $T^{+} = 15$ or $T^{+} = 0$). Six of nine comparisons reach this floor, confirming a consistent directional effect (classical $>$ quantum) that nonetheless cannot reach $\alpha = 0.05$ significance with only five folds.

\subsection{Friedman Tests Within Datasets}
\label{app:friedman}

\begin{table}[htbp]
\centering
\caption{Friedman test results within each dataset (all quantum + classical configs). All datasets show significant within-method variation ($p < 0.001$).}
\label{tab:friedman_full}
\small
\begin{tabular}{@{}lrrrl@{}}
\toprule
Dataset & $\chi^2$ & df & $p$-value & Significant? \\
\midrule
banknote       & 128.6 & 27 & $< 0.001$ & Yes \\
breast\_cancer & 219.6 & 47 & $< 0.001$ & Yes \\
diabetes\_pima & 170.5 & 47 & $< 0.001$ & Yes \\
haberman       &  72.7 & 27 & $< 0.001$ & Yes \\
heart\_disease & 129.1 & 35 & $< 0.001$ & Yes \\
ionosphere     & 206.2 & 47 & $< 0.001$ & Yes \\
parkinson\_489 &  83.3 & 35 & $< 0.001$ & Yes \\
sonar          & 184.1 & 47 & $< 0.001$ & Yes \\
spambase       &  59.8 & 15 & $< 0.001$ & Yes \\
\botrule
\end{tabular}
\end{table}

\noindent Interpretation: while methods significantly differ within each dataset, the best quantum configuration never statistically outperforms the best classical baseline in pairwise tests.

\subsection{Kernel Spectral Properties}
\label{app:spectral_full}

\begin{table}[htbp]
\centering
\caption{Representative kernel eigenspectrum ranges (extended-study ideal/classical kernel matrices, $n_{\text{train}} \approx 400$--$1{,}300$). Ranges correspond to the most common configurations; extreme-$k$ or very small datasets may produce values outside these intervals.}
\label{tab:spectral_full}
\small
\begin{tabular}{@{}lccccr@{}}
\toprule
Kernel & Eff.\ Rank Ratio & Top-1 Expl. & Top-5 Expl. & Diag.\ Dom. & Neg.\ $\lambda$ \\
\midrule
belis (quantum)   & 0.40--0.74 & 0.03--0.06 & 0.07--0.13 & 18--39 & 0\% \\
rot2dof (quantum) & 0.01--0.02 & 0.52--0.57 & 0.66--0.72 & 1.9--2.2 & 0\% \\
linear (classical)  & $\sim$0.01 & 0.27--0.29 & 0.76--0.80 & 568 & ${\leq}0.4$\% \\
poly3 (classical)   & $\sim$0.03 & 0.22--0.26 & 0.62--0.63 & 154--168 & ${\leq}1.3$\% \\
\textbf{rbf (classical)} & \textbf{0.06--0.07} & \textbf{0.31--0.35} & \textbf{0.52--0.54} & \textbf{3.7--3.8} & \textbf{0\%} \\
\botrule
\end{tabular}
\end{table}

\noindent The ``Goldilocks zone'' is visible in the effective rank ratio: belis is too high (near-identity), rot2dof is too low (near-rank-1), while RBF sits at an intermediate value that provides the best classification performance.

\subsection{Computational Cost}
\label{app:cost_full}

Table~\ref{tab:cost_full} reports the mean wall-clock time for computing a single kernel matrix (training split) across all main-benchmark and extended-study experiments, measured on a single CPU core with statevector simulation. Classical kernels (linear, RBF, polynomial) are computed via scikit-learn in $\sim$0.3\,s. Quantum kernels require $\sim$3\,s ($10\times$) due to the overhead of simulating the fidelity circuit for each sample pair. Quantum kernel training (QKT) amplifies cost further, as each KTA optimisation step requires a full kernel recomputation; with 50--170 gradient iterations per fold, the aggregate overhead reaches $\sim$2\,000$\times$. These timings exclude hardware queue latency, which adds additional overhead for on-device execution (Appendix~\ref{app:hardware}).

\begin{table}[htbp]
\centering
\caption{Mean kernel computation time by category and cost ratios relative to classical.}
\label{tab:cost_full}
\small
\begin{tabular}{@{}lrr@{}}
\toprule
Category & Mean Kernel Time (s) & Ratio vs.\ Classical \\
\midrule
Classical (linear/rbf/poly) & 0.31 & $1\times$ \\
Quantum (belis/rot2dof/\ldots) & 3.07 & $10\times$ \\
QKT (kernel training) & 433--837 & $1{,}400$--$2{,}700\times$ \\
\botrule
\end{tabular}
\end{table}

\subsection{Circuit Depth Scaling}
\label{app:circuit_depth}

Table~\ref{tab:circuit_depths} summarises pre-transpilation circuit depths and two-qubit gate counts for the four quantum feature maps at representative feature dimensions ($k \in \{4, 8, 14\}$), all with $R = 2$ repetitions. Rot2DoF maintains constant depth 12 regardless of $k$ because it contains no entangling gates. Belis and Sakhnenko10 scale linearly with the number of qubits due to their CNOT structures. ZZFeatureMap reports a nominal depth of 1 at the parameterised-circuit level; its actual depth after transpilation to a hardware gate set depends on the target topology. Detailed transpiled properties on ibm\_fez are given in Appendix~\ref{app:feature_maps}.

\begin{table}[htbp]
\centering
\caption{Feature map circuit depths and gate counts at $R{=}2$ repetitions.}
\label{tab:circuit_depths}
\small
\begin{tabular}{@{}lccccl@{}}
\toprule
Feature Map & $k{=}4$ & $k{=}8$ & $k{=}14$ & CX ($k{=}8$) & Entanglement \\
\midrule
rot2dof        & 12 & 12 & 12 & 0  & No \\
belis          & 26 & 29 & 32 & 14 & Yes (linear) \\
sakhnenko10    & 44 & 48 & 54 & 16 & Yes (ring) \\
zzfm           & 31 & 67 & 121 & 24/112/364$^*$ & Yes ($ZZ$) \\
\botrule
\end{tabular}
\end{table}
\noindent $^*$CX counts at $k{=}4$/$k{=}8$/$k{=}14$ respectively (post-decomposition; see also Table~\ref{tab:feature_maps}).

\subsection{Learning Curve Slopes}
\label{app:learning_slopes}

Table~\ref{tab:learning_slopes_full} extends the main-text learning curve analysis (Table~\ref{tab:slopes}) by reporting per-category slopes and their statistical significance. Slopes are computed as the ordinary least-squares coefficient of balanced accuracy regressed on $\log n_{\text{train}}$ across six training fractions (10\%, 20\%, 30\%, 50\%, 70\%, 100\%). Significance is assessed via a two-sided $t$-test on the slope coefficient at $\alpha = 0.05$. Parkinson\_489 is excluded because it was added in Phase~3 under a different pipeline configuration.

Quantum ideal slopes are significantly positive on all eight datasets, whereas classical slopes fail to reach significance on two (diabetes\_pima, haberman). This confirms that quantum kernels exhibit steeper learning rates on the majority of datasets, despite starting from a lower absolute baseline (see Table~\ref{tab:gap_evolution_full}). Noisy simulation slopes closely track their ideal counterparts, indicating that depolarising noise does not substantially alter the data-efficiency profile.

\begin{table}[htbp]
\centering
\caption{Learning curve slopes (BA vs.\ $\log n_{\text{train}}$) with significance. Positive slopes indicate performance improves with more training data. Parkinson\_489 is excluded as it was added in Phase~3 with a different pipeline configuration.}
\label{tab:learning_slopes_full}
\small
\begin{tabular}{@{}l cc cc cc@{}}
\toprule
& \multicolumn{2}{c}{Classical} & \multicolumn{2}{c}{Q.\ Ideal} & \multicolumn{2}{c}{Q.\ Noisy} \\
\cmidrule(lr){2-3} \cmidrule(lr){4-5} \cmidrule(lr){6-7}
Dataset & Slope & Sig. & Slope & Sig. & Slope & Sig. \\
\midrule
banknote       & 0.009 & Yes & 0.015 & Yes & 0.009 & No  \\
breast\_cancer & 0.024 & Yes & 0.032 & Yes & 0.032 & Yes \\
diabetes\_pima & 0.008 & No  & 0.023 & Yes & 0.008 & No  \\
haberman       & 0.010 & No  & 0.026 & Yes & 0.024 & Yes \\
heart\_disease & 0.063 & Yes & 0.024 & Yes & 0.029 & Yes \\
ionosphere     & 0.069 & Yes & 0.047 & Yes & 0.047 & Yes \\
sonar          & 0.068 & Yes & 0.079 & Yes & 0.084 & Yes \\
spambase       & 0.007 & Yes & 0.011 & Yes & 0.012 & Yes \\
\botrule
\end{tabular}
\end{table}

\subsection{Quantum vs.\ Classical Gap Evolution}
\label{app:gap_evolution}

Table~\ref{tab:gap_evolution_full} tracks the performance gap between the best quantum ideal and best classical configuration at two training fractions: 10\% and 100\% of the available training data. Negative values indicate classical superiority. The ``Narrows?'' column indicates whether the gap decreases (in absolute terms) from 10\% to 100\% training data; ``Q~Surpasses?'' indicates whether the quantum configuration achieves higher BA at any fraction.

On six of eight datasets, the gap narrows as training size increases, consistent with the steeper quantum learning slopes reported in Table~\ref{tab:learning_slopes_full}. However, only haberman sees a full crossover: the gap moves from $-0.002$ at 10\% to $+0.046$ at 100\%, making it the only dataset where the quantum kernel eventually outperforms the classical representative. Heart\_disease and ionosphere show an initial marginal quantum advantage at 10\% training data that reverses at full data, suggesting that the quantum feature space saturates earlier than the classical one on these datasets.

\begin{table}[htbp]
\centering
\caption{Quantum--classical BA gap at 10\% and 100\% training fraction (negative = classical better).}
\label{tab:gap_evolution_full}
\small
\begin{tabular}{@{}lrrll@{}}
\toprule
Dataset & Gap@10\% & Gap@100\% & Narrows? & Q Surpasses? \\
\midrule
banknote       & $-0.086$ & $-0.077$ & Yes & No  \\
breast\_cancer & $-0.079$ & $-0.063$ & Yes & No  \\
diabetes\_pima & $-0.098$ & $-0.062$ & Yes & No  \\
haberman       & $-0.002$ & $+0.046$ & Yes & \textbf{Yes} \\
heart\_disease & $+0.008$ & $-0.110$ & No  & Yes* \\
ionosphere     & $+0.012$ & $-0.047$ & No  & Yes* \\
sonar          & $-0.139$ & $-0.084$ & Yes & No  \\
spambase       & $-0.032$ & $-0.025$ & Yes & No  \\
\botrule
\end{tabular}
\end{table}
\noindent *At 10\% training fraction only; advantage reverses at full data.

\subsection{Seed Sensitivity: Paired Analysis}
\label{app:seed_full}

\begin{table}[htbp]
\centering
\caption{Seed stability (CoV of mean BA across 16 seeds) and paired quantum vs.\ classical analysis. Parkinson\_489 and spambase are excluded: seed analysis was conducted on the seven Phase~1 datasets.}
\label{tab:seed_full}
\small
\begin{tabular}{@{}l ccc crrl@{}}
\toprule
& \multicolumn{3}{c}{CoV (\%)} & & \multicolumn{3}{c}{Paired Analysis} \\
\cmidrule(lr){2-4} \cmidrule(lr){6-8}
Dataset & Class. & Q.I. & Q.N. & & QI Wins & Adv. & $p$ \\
\midrule
banknote       & 0.02 & 0.24 & 0.41 & & 0/16 & $-0.084$ & $<0.001$ \\
breast\_cancer & 0.46 & 0.70 & 0.66 & & 0/16 & $-0.069$ & $<0.001$ \\
diabetes\_pima & 0.75 & 1.74 & 1.59 & & 0/16 & $-0.053$ & $<0.001$ \\
haberman       & 3.30 & 2.60 & 2.44 & & 14/16 & $+0.019$ & 0.004 \\
heart\_disease & 1.89 & 1.81 & 1.41 & & 0/16 & $-0.072$ & $<0.001$ \\
ionosphere     & 0.81 & 0.84 & 0.85 & & 0/16 & $-0.055$ & $<0.001$ \\
sonar          & 1.81 & 2.81 & 2.37 & & 0/16 & $-0.081$ & $<0.001$ \\
\botrule
\end{tabular}
\end{table}

\noindent Haberman is the only dataset where quantum ideal consistently outperforms classical (14/16 seeds, $p = 0.004$). On all other datasets, classical wins on every seed.

\subsection{QKT Theta Stability}
\label{app:qkt_stability}

\begin{table}[htbp]
\centering
\caption{QKT optimisation stability summary across 28 configurations.}
\label{tab:qkt_stability}
\small
\begin{tabular}{@{}lp{0.45\columnwidth}@{}}
\toprule
Metric & Value \\
\midrule
Configurations analysed & 28 \\
Mean theta CoV across folds & 0.535 \\
Median pairwise rank corr. & 0.738 \\
Mean KTA improvement (final $-$ initial) & $+0.130$ \\
Convergence rate (within tol.) & 13.6\% \\
Best QKT result & BA $= 0.968$ (breast\_cancer, tree-k8-belis) \\
\hspace{1.5em}Theta CoV & 0.050 \\
\hspace{1.5em}Rank correlation & 0.973 \\
\botrule
\end{tabular}
\end{table}

\noindent The best QKT configuration (breast\_cancer, tree-k8-belis) is also the most stable ($\CoV = 0.050$, rank correlation $= 0.973$), suggesting that QKT effectiveness and parameter stability are correlated.

\subsection{Dataset Suitability Criterion}
\label{app:suitability}

\begin{table}[htbp]
\centering
\caption{Dataset suitability for quantum kernels. $N$: post-cleaning sample count (identical to Table~\ref{tab:datasets}). Non-linearity gap $= \BA_{\text{rbf}} - \BA_{\text{linear}}$; $\Delta = \BA_{\text{Q}} - \BA_{\text{C}}$.}
\label{tab:suitability_full}
\small
\begin{tabular}{@{}lrrrrrl@{}}
\toprule
Dataset & $N$ & BA\textsubscript{C} & BA\textsubscript{Q} & $\Delta$ & NL Gap & Favorable? \\
\midrule
haberman       & 306   & 0.549 & 0.581 & $+0.032$ & $+0.043$ & \textbf{Yes} \\
spambase       & \num{4601} & 0.903 & 0.888 & $-0.016$ & $+0.017$ & No \\
ionosphere     & 351   & 0.935 & 0.905 & $-0.031$ & $+0.079$ & No \\
sonar          & 208   & 0.872 & 0.813 & $-0.059$ & $+0.086$ & No \\
diabetes\_pima & 768   & 0.729 & 0.669 & $-0.060$ & $-0.008$ & No \\
breast\_cancer & 569   & 0.976 & 0.913 & $-0.063$ & $+0.006$ & No \\
banknote       & \num{1372} & 1.000 & 0.921 & $-0.079$ & $+0.010$ & No \\
parkinson\_489 & 489   & 0.829 & 0.746 & $-0.083$ & $+0.017$ & No \\
heart\_disease & 303   & 0.844 & 0.724 & $-0.120$ & $-0.008$ & No \\
\botrule
\end{tabular}
\end{table}

\noindent Spearman's $\rho = 0.683$ ($p = 0.042$) between non-linearity gap and $\Delta$, suggesting that datasets where RBF already provides large gains over linear are not the datasets where quantum kernels help -- a potentially useful pre-screening criterion.

\subsection{Computational Cost Comparison}
\label{app:cost_comparison}

\begin{figure}[htbp]
    \centering
    \includegraphics[width=0.65\columnwidth]{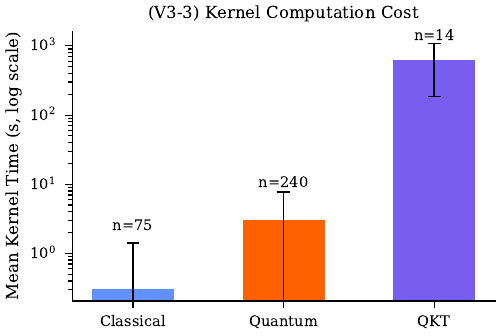}
    \caption{Computational cost comparison (log scale). Classical kernels complete in $\sim$0.31\,s, quantum statevector simulation requires $\sim$3.1\,s ($10\times$), and quantum kernel training (QKT) requires $\sim$635\,s ($2{,}060\times$).}
    \label{fig:cost_comparison}
\end{figure}

Figure~\ref{fig:cost_comparison} compares wall-clock times for classical kernel computation, quantum statevector simulation, and quantum kernel training (QKT) with iterative KTA optimisation. Discussion is in Sect.~\ref{sec:results_statistical}.

\subsection{Learning Curves and Slope Analysis}
\label{app:learning_curves_main}

\begin{figure}[htbp]
    \centering
    \includegraphics[width=\columnwidth]{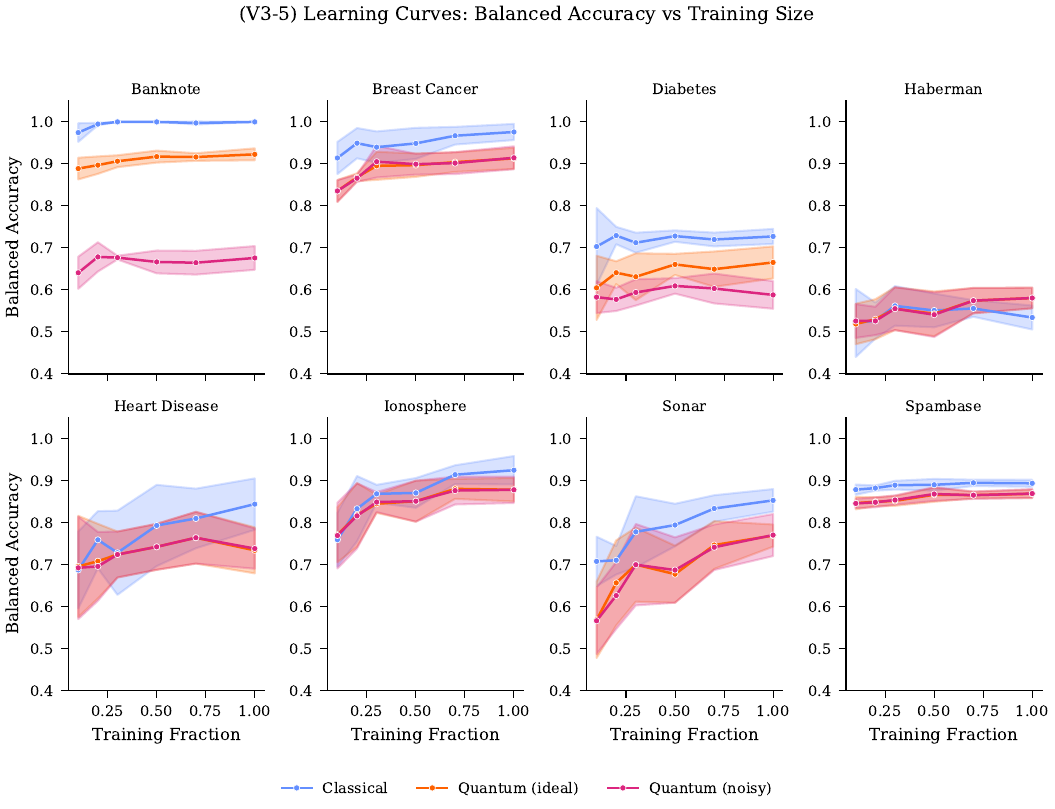}
    \caption{Learning curves for all eight datasets across six training fractions (10\%--100\%). Each panel shows classical (blue), quantum ideal (orange), and quantum noisy (magenta) balanced accuracy as a function of training set size. Quantum kernels show steeper slopes on 6/8 datasets but fail to close the gap.}
    \label{fig:learning_curves}
\end{figure}

Figure~\ref{fig:learning_curves} shows per-dataset learning curves at six training fractions, and Table~\ref{tab:slopes} reports the corresponding OLS slopes and gap evolution. Discussion is in Sect.~\ref{sec:results_learning}.

\begin{table}[htbp]
\caption{Learning curve slopes (BA vs.\ $\log n_{\text{train}}$) and gap evolution. Slopes are computed via OLS regression across six training fractions. Quantum ideal slopes are steeper on 6/8 datasets. ``Gap narrows'' indicates the quantum--classical gap decreases from 10\% to 100\% training data.}
\label{tab:slopes}
\centering
\footnotesize
\begin{tabular}{@{}lccccc@{}}
\toprule
\textbf{Dataset} & \textbf{Classical} & \textbf{Q-Ideal} & \textbf{Q-Noisy} & \textbf{Gap} & \textbf{Gap}\\
 & \textbf{Slope} & \textbf{Slope} & \textbf{Slope} & \textbf{@10\%} & \textbf{@100\%}\\
\midrule
banknote        & 0.009 & \textbf{0.015} & 0.009 & $-$8.6 & $-$7.7\\
breast\_cancer  & 0.024 & \textbf{0.032} & \textbf{0.032} & $-$7.9 & $-$6.3\\
diabetes\_pima  & 0.008 & \textbf{0.023} & 0.008 & $-$9.8 & $-$6.2\\
haberman        & 0.010 & \textbf{0.026} & \textbf{0.024} & $-$0.2 & $\mathbf{+4.6}$\\
heart\_disease  & \textbf{0.063} & 0.024 & 0.029 & $+$0.8 & $-$11.0\\
ionosphere      & \textbf{0.069} & 0.047 & 0.047 & $+$1.2 & $-$4.7\\
sonar           & 0.068 & \textbf{0.079} & \textbf{0.084} & $-$13.9 & $-$8.4\\
spambase        & 0.007 & \textbf{0.011} & \textbf{0.012} & $-$3.2 & $-$2.5\\
\botrule
\end{tabular}
\end{table}


\FloatBarrier

\section*{Supplementary Information}

The online supplementary material includes all 39 figures, per-fold balanced accuracy arrays for all 970 experiments, and the complete set of 683 cached kernel matrices in NumPy compressed format.

\section*{Acknowledgements}

This work has been supported by the Schaeffler Hub for Advanced Research at the Friedrich-Alexander-Universit\"at Erlangen-N\"urnberg (SHARE at FAU). The authors thank IBM Quantum for providing access to the ibm\_fez quantum processor through the IBM Quantum Open Plan. Computational experiments were performed using the Qiskit~2.1+ framework.

\section*{Statements and Declarations}

\paragraph{Funding.}
This research was funded by Schaeffler Technologies AG \& Co.\ KG.

\paragraph{Competing interests.}
The authors declare no competing interests.

\paragraph{Data availability.}
All datasets used in this study are publicly available from the UCI Machine Learning Repository~\citep{uci2017} and OpenML~\citep{vanschoren2014openml}. The complete benchmark suite, including cached kernel matrices, hardware results, and figure generation scripts, is archived at \url{https://doi.org/10.5281/zenodo.19197916}.

\paragraph{Code availability.}
The complete benchmark framework ($\sim$18,700 lines of Python code) is open-source and publicly archived at \url{https://doi.org/10.5281/zenodo.19197916}, including 132 unit tests and all experiment runner scripts.

\paragraph{Use of AI-assisted technologies.}
The entire codebase and manuscript were independently developed, implemented, and written by the authors. AI-assisted tools were used solely for refining readability, optimising code quality, and proof-reading. All AI-generated suggestions were critically reviewed and verified by the authors before incorporation. The authors take full responsibility for the content of this publication.

\paragraph{Author contribution.}
S.K.\ conceived and designed the study, implemented all software, conducted all experiments (including hardware execution on IBM ibm\_fez), performed the analysis, generated all figures, and wrote the manuscript. C.S.\ and M.S.\ contributed to manuscript review, proofreading, and technical verification. All authors reviewed and approved the final manuscript.


\bibliographystyle{plainnat}
\bibliography{references}

\end{document}